\documentclass[fleqn,usenatbib]{mnras}
\usepackage{multirow}
\usepackage{graphicx}
\usepackage{times}
\usepackage{float}
\usepackage{pdflscape}
\usepackage{threeparttable}
\usepackage{amsmath}
\usepackage{amssymb}
\usepackage{xspace}
\usepackage{newtxtext}
\usepackage{color}
\title[Broadband X-ray emission in Ton\,S180]{The first broadband X-ray view of the narrow line Seyfert\,1 Ton\,S180}
\author[Matzeu et al.]
{G. A. Matzeu$^{1,2}$\thanks{Correspondence to: gabriele.matzeu@sciops.esa.int},
 E. Nardini$^{3,4}$, M. L. Parker$^{1}$, J. N. Reeves$^{5}$, V. Braito$^{2,5}$,   
 \newauthor
D. Porquet$^{6}$, R. Middei$^{7,8}$, E. Kammoun$^{9}$, E. Lusso$^{3,4}$, W. N. Alston$^{10}$, M. Giustini$^{11}$,   
\newauthor
A. P. Lobban$^{1}$, A. M. Joyce$^{1}$, Z. Igo$^{1}$, J. Ebrero$^{1}$, L. Ballo$^{1}$, M. Santos-Lle{\'o}$^{1}$,
\newauthor
and N. Schartel$^1$\\
$^{1}$European Space Agency (ESA), European Space Astronomy Centre (ESAC), E-28691 Villanueva de la Ca\~{n}ada, Madrid, Spain\\
$^{2}$INAF -- Osservatorio Astronomico di Brera, Via Bianchi 46, I-23807 Merate (LC), Italy\\
$^3$Dipartimento di Fisica e Astronomia, Universit{\` a} di Firenze, via G. Sansone 1, 50019 Sesto Fiorentino, Firenze, Italy
\\
$^4$INAF -- Osservatorio Astrofisico di Arcetri, Largo Enrico Fermi 5, I-50125 Firenze, Italy\\
$^5$Department of Physics, Institute for Astrophysics and Computational Sciences, The Catholic University of America, Washington, DC\,20064, USA \\
$^6$Aix-Marseille Univ, CNRS, CNES, LAM, Marseille, France\\
$^7$INAF - Osservatorio Astronomico di Roma, Via Frascati 33, 00078, Monte Porzio Catone (Roma), Italy\\
$^8$Space Science Data Center - ASI, Via del Politecnico s.n.c., 00133 Roma, Italy\\
$^9$Department of Astronomy, University of Michigan, 1085 South University Avenue, Ann Arbor, MI 48109-1107, USA\\
$^{10}$Institute of Astronomy, Madingley Rd, Cambridge, CB3 0HA\\
$^{11}$Centro de Astrobiolog\'ia (CSIC-INTA), Camino Bajo del Castillo s/n,  
Villanueva de la Ca\~nada, E-28692 Madrid, Spain\\}

		\newcommand{\Msun}{\mbox{$M_{\odot}$}\xspace}

		\newcommand{\ang}{\mathring{\mathrm{A}}}
 		
		\newcommand{\feka}{Fe\,K$\alpha$\xspace}

		\newcommand{\vout}{v_{\rm out}\xspace}

		\newcommand{\dchidof}{\Delta\chi^{2}/\Delta\nu}
		\newcommand{\chidof}{\chi^{2}/\nu}
		\newcommand{\chisq}{\chi^{2}}
		\newcommand{\nhgal}{N_{\rm H}^{\rm Gal}}
		\newcommand{\nh}{N_{\rm H}}
		
		\newcommand{\logxi}{\log(\xi/\rm{erg\,cm\,s}^{-1})}

		\newcommand{\lbol}{L_{\rm bol}}

		\newcommand{\lion}{L_{\rm ion}}
		\newcommand{\ledd}{L_{\rm Edd}}
		
		\newcommand{\mbh}{M_{\rm BH}}

		\newcommand{\fe}{Fe\,K\xspace}
        \newcommand{\iron}{iron\,K\xspace}
		\newcommand{\ergs}{\rm erg\,s^{-1}\xspace}
		\newcommand{\cmsq}{{\rm cm$^{-2}$}\xspace}
		\newcommand{\cmq}{{\rm cm$^{-3}$}\xspace}
		
		\newcommand{\rg}{r_{\rm g}}
		\newcommand{\ev}{\,\rm eV}		
		\newcommand{\kev}{\,\rm keV}

		\newcommand{\kms}{\rm km\,s^{-1}\xspace}

		\newcommand{\fig}{Fig.\,}

		\newcommand{\ks}{\rm ks}
		\newcommand{\chis}{\chi^{2}/\nu}

		\newcommand{\leddratio}{L_{\rm bol}/L_{\rm Edd}}
		
		\newcommand{\logMdot}{\dot M/\dot{M}_{\rm Edd}}
		\newcommand{\logmdot}{\log(\dot m)}

		\newcommand{\suzaku}{\emph{Suzaku}\xspace} 
		\newcommand{\nustar}{\emph{NuSTAR}\xspace}		
		\newcommand{\xmm}{\emph{XMM-Newton}\xspace}

		\newcommand{\xmmnu}{\textit{XMM-Newton} and \textit{NuSTAR}\xspace}

		\newcommand{\pds}{PDS\,456\xspace}		
				
		\newcommand{\ton}{Ton\,S180\xspace}		
		\newcommand{\arkone}{Ark\,120\xspace}

		\newcommand{\xillver}{\texttt{xillver}\xspace}
		\newcommand{\relxill}{\texttt{relxill}\xspace}
		
                \newcommand{\relxilllpion}{\texttt{relxilllpion}\xspace}
		\newcommand{\relxillcp}{\texttt{relxillCp}\xspace}
		\newcommand{\relxilld}{\texttt{relxillD}\xspace}

		\newcommand{\optxagn}{\texttt{optxagnf}\xspace}
		\newcommand{\optxconv}{\texttt{optxconv}\xspace}
		\newcommand{\nthcomp}{\texttt{nthComp}\xspace}
		\newcommand{\xspecmcmc}{\texttt{xspec$\textunderscore$emcee}\xspace}
		\newcommand{\agnslim}{\texttt{agnslim}\xspace}
		\newcommand{\heasoft}{\textsc{HEAsoft}\xspace}
		\newcommand{\xstar}{\textsc{xstar}\xspace}
		\newcommand{\xspec}{\textsc{xspec}\xspace}

		\newcommand{\rgsproc}{\texttt{rgsproc}\xspace}
		\newcommand{\sas}{\textsc{sas}\xspace}
		\newcommand{\ftools}{\textsc{ftools}\xspace}
		\newcommand{\rgscombine}{\texttt{rgscombine}\xspace}
		
		\newcommand{\nupip}{\texttt{nupipeline}\xspace}
		\newcommand{\nusaa}{\texttt{nucalcsaa}\xspace}
		\newcommand{\omi}{\texttt{omichain}\xspace}

\begin{document}

\date{\today}

\pagerange{\pageref{firstpage}--\pageref{}} \pubyear{?}

\maketitle
\label{firstpage}

\maketitle
\begin{abstract}

	We present joint \textit{XMM-Newton} and \textit{NuSTAR} observations of the `bare' narrow line Seyfert 1 \ton ($z=0.062$), carried out in 2016 and providing the first hard X-ray view of this luminous galaxy. We find that the 0.4--30 keV band cannot be self-consistently reproduced by relativistic reflection models, which fail to account simultaneously for the soft and hard X-ray emission. The smooth soft excess prefers extreme blurring parameters, confirmed by the nearly featureless nature of the RGS spectrum, while the moderately broad \fe line and the modest hard excess above 10 keV appear to arise in a milder gravity regime. By allowing a different origin of the soft excess, the broadband X-ray spectrum and overall spectral energy distribution (SED) are well explained by a combination of: (a) direct thermal emission from the accretion disc, dominating from the optical to the far/extreme UV; (b) Comptonization of seed disc photons by a warm ($kT_{\rm e}\sim0.3$ keV) and optically thick ($\tau\sim10$) corona, mostly contributing to the soft X-rays; (c) Comptonization by a standard hot ($kT_{\rm e}\ga100$ keV) and optically thin ($\tau<0.5$) corona, responsible for the primary X-ray continuum; and (d) reflection from the mid/outer part of the disc. The two coronae are suggested to be rather compact, with $R_{\rm hot}\la R_{\rm warm}\la 10\,\rg$. Our SED analysis implies that \ton accretes at super-Eddington rates. This is a key condition for the launch of a wind, marginal (i.e., 3.1$\sigma$ significance) evidence of which is indeed found in the RGS spectrum.

\end{abstract}

\section{introduction}

It is now common knowledge that in most Seyfert AGN, the primary UV to X-ray emission can be attributed to complex interactions between the accretion disc and a corona of relativistic electrons. In the standard picture the `seed' optical/UV disc photons that are Compton up-scattered in the hot corona \citep{HaardtMaraschi91,HaardtMaraschi93} can be observed as a hard X-ray tail, which is usually described phenomenologically by a simple power law up to $\sim$10 keV. However, more complex emission features are also imprinted on the X-ray spectra of AGN, namely the soft excess below $\sim$1--2 keV, a \feka line complex around 6.4 keV and a Compton `hump' peaking at $\sim$20 keV. 

The soft excess is a smooth (i.e., featureless) emission component that is commonly observed in unabsorbed AGN, where the power law fails to account for this extra emission. Studies conducted by \citet{Gierliski04}, \citet{Porquet04}, \citet{Piconcelli05} and \citet{Miniutti09} have shown that the soft excess cannot be directly associated to the Wien tail of the the blackbody-like emission from accretion disc observed in the UV, as previously thought \citep[e.g.,][]{Singh85,Pounds86,WalterFink93}, since the thermal continuum requires temperatures far higher than expected. 

\citet{Done12} suggested that an increase of the disc effective temperature, caused by Comptonization of the `seed' photons in a cold ($kT_{\rm e} < 1$ keV) and optically thick plasma $(\tau\gg1)$, may be the reason for the observed excess. \citet{Petrucci13} carried out a detailed analysis on the data obtained from a large multiwavelength \xmm and \textit{INTEGRAL} campaign on the bright Seyfert\,1 AGN Mrk\,509 \citep{Kaastra11}. The broadband (i.e., optical/UV to hard X-rays) spectrum of Mrk\,509 was well described with the contribution from: (i) a hot ($kT_{\rm e}\sim100$ keV) and optically thin ($\tau\sim0.5$) corona responsible for the primary continuum and (ii) a warm ($kT_{\rm e}\sim1$ keV) and optically thick ($\tau\sim10$--$20$) plasma for the soft X-ray component (i.e., the soft excess). The differences between \citet{Done12} and \citet{Petrucci13} are both geometrical and physical. Both scenarios require a warm corona with comparable optical depth and temperature but located in different places with respect to the hot corona, respectively within or beyond a given characteristic radius. Moreover, in the latter case, the disc is entirely `passive', since all the accretion power is released in the warm corona making the disc intrinsic emission negligible.\footnote{In this framework, the accretion power is not dissipated within the disc giving rise to the typical accretion disc spectrum \citep[e.g.,][]{Shakura73}, but exclusively in the warm corona. A visual illustration of the implications in terms of the emitted UV spectrum is later shown in Fig.\,\ref{fig:ts180_2016_2nth_eeuf_ra.pdf}.} 

This physical interpretation is referred to as the \textit{two-corona model}, and it has been successfully tested on several local Seyfert galaxies, among which Mrk\,509 \citep{Petrucci13}, RX\,J0439.6--5311 \citep{Jin17}, Ark\,120 \citep{Porquet18}, NGC\,7469 \citep{Middei18}, NGC\,4593 \citep{Middei19}, and HE\,1143--1810 \citep{Ursini19arXiv}. In addition, \citet{Matzeu17} suggested that the interplay between a dual (warm and hot) coronal region could be the cause of the intrinsic short-term spectral variability in the quasar \pds, caught in a high-flux and unabsorbed state by \suzaku in 2007.

Another viable explanation is that the soft X-ray excess is due to reflection from the photoionized surface layers in the inner region (near the supermassive black hole) of the accretion disc. In this strong-gravity regime the extreme relativistic blurring reduces the (narrow) fluorescence soft X-ray emission lines into a featureless continuum \citep{Fabian02,Ross05,Crummy06,Garcia10,Nardini11,Lohfink12,Walton13,WilkinsGallo15,Jiang19}. High ionization also contributes to lessening the prominence of the reflection features against the direct continuum. Sometimes this scenario requires extreme solutions, such as very high values of black hole spin parameter\footnote{The spin parameter $a$ is defined as $cJ/GM^2$, where $J$ and $M$ are the black hole's angular momentum and mass, respectively. For maximal spin, $a=0.998$, the innermost stable circular orbit (ISCO) reduces to $R_{\rm isco}\simeq 1.24$ gravitational radii ($\rg=GM/c^2$), as opposed to $R_{\rm isco}=6\,\rg$ for a non-spinning Schwarzschild black hole, with $a = 0$ \citep{Thorne74}.}, i.e. $ a \to 0.998$ \citep[see][]{Reynolds14}, and of the disc emissivity index\footnote{The disc emissivity is typically assumed to follow a power-law dependence on radius, $\epsilon \propto r^{-q}$. For a point source in a flat Euclidean space, it is $\epsilon \propto r^{-3}$ at large distance.}, $q \to 10$. However, a variety of reflection models have been adopted to successfully account for the soft X-ray excess; in particular, a considerable improvement was obtained by implementing a high-density accretion disc, up to $\log( n/\rm cm^{-3})=19$ \citep{Garcia16}. Regardless of the validity of all the above interpretations, the physical origin of the soft X-ray excess component is still an open issue after many years of AGN research.

Tonantzintla (Ton) S180 is a local \citep[$z=0.06198$; ][]{Wisotzki95}, luminous \citep[$\lbol\sim5\times10^{45}\,\ergs$; ][]{Turner02} narrow line Seyfert~1 (NLSy1), which is considered one of the prototypical `bare' AGN with no trace of absorption and a featureless and prominent soft excess \citep{Turner01,Vaughan02b}. The 2007 \textit{Suzaku} observation (102 ks) suggested also an intriguing hard excess in the \textit{HXD/PIN} energy range \citep[$E\sim15$--$55$ keV;][]{Takahashi10}. However, since the \textit{HXD} was a non-focusing detector \citep{Takahashi07}, proper background subtraction is very critical especially for faint sources at $E>10\kev$ as \ton. 
\citet{Nardini12} presented a spectral analysis of the \suzaku and \xmm observations of \ton. It was found that a self-consistent dual-reflector geometry reproduced effectively the main X-ray spectral properties, namely the soft excess, the broad \fe emission feature and the hard X-ray emission (tentatively up to $\sim$30 keV). \ton was targeted three times by \xmm in 2000, 2002 and 2015. For the latter observation, \citet{Parker18ton} found that the \xmm spectrum favoured two Comptonization components plus reflection from a disc around a black hole of low spin (with maximal spin ruled out at the $3\sigma$ confidence level).

In this work we present a detailed analysis of the 2016 joint \xmmnu observation (with durations of $\sim$30 and 270 ks, respectively) of \ton (PI: G.A. Matzeu). Here, for the first time, the hard X-ray spectrum of \ton above $10\,\kev$ from a direct-imaging hard X-ray telescope is revealed. The main goal of this paper is to test whether the soft excess and the broadband \xmmnu 2016 spectra can be explained with either relativistic reflection or a warm corona scenario. Both interpretations have been proposed in the past but the lack of high-quality data above 10 keV prevented any conclusive resolution. 

This paper is organized as follows: in Section\,\ref{sec:Observations and Data Reduction} we summarize the data reduction process for each of the \xmm and \nustar detectors, whereas in Section\,\ref{subsec:The broadband xmmnu analysis} we present the Reflection Grating Spectrometer (RGS), X-ray broadband and optical to X-ray spectral energy distribution (SED) analysis, where we test relativistic reflection models as well as two-corona/multi-temperature Comptonized accretion disc models. In Section\,\ref{sec:Discussion}, the physical implications from the above analysis are discussed in detail.

In this paper the values of $H_0=70$\,km\,s$^{-1}$\,Mpc$^{-1}$, $\Omega_{\Lambda_{0}}=0.73$ and $\Omega_{\rm M}=0.27$ are assumed throughout, and errors are quoted at the 90 per cent confidence level ($\Delta \chi^{2}=2.71$) for one parameter of interest, unless otherwise stated. 
  
%\begin{landscape}
\begin{table*}
%\begin{minipage}{100mm}
\centering
%\scriptsize
\begin{tabular}{c@{\hspace{25pt}}c@{\hspace{15pt}}c@{\hspace{15pt}}c@{\hspace{15pt}}c@{\hspace{15pt}}c}
\hline \\ [-2ex]

%Obs.~ID &  Instrument     &&&&&&\\

%                                                    &\multicolumn{3}{c}{\xmm}                                                                         &\multicolumn{2}{c}{\nustar}\\ 

Observation                        &\multicolumn{5}{c}{$2016$}\\[2ex]

Telescope                          &\multicolumn{3}{c@{\hspace{23pt}}}{\xmm}                                                           &\multicolumn{2}{c@{\hspace{13pt}}}{\nustar}\\[2ex]

Obs.~ID  &\multicolumn{3}{c@{\hspace{22pt}}}{$0790990101$} &\multicolumn{2}{c@{\hspace{14pt}}}{$60101027002$}\\[2ex]

Instrument      &EPIC-pn    &EPIC-MOS\,1$+$2        &RGS\,1$+$2    &FPMA         &FPMB\\[2ex]           

Start Date             &2016--06--13  &2016--06--13  &2016--06--13  &2016--06--10   &2016--06--10\\[1ex]

 Time\,(UT)            &05:42:30	   &05:36:42  &05:36:33		&12:31:08     &12:31:08\\[2ex]

End Date       &2016--06--13  &2016--06--13  &2016--06--13  &2016--06--13  &2016--06--13\\[1ex]
Time\,(UT)      &14:10:27	 &14:07:43  &14:11:45	 &15:16:08    &15:16:08\\[2ex]

Duration\,(ks)            &30.5      &61.3    &30.9        &268.6         &268.6\\[2ex]

Exposure\,(ks)$^{\rm a}$  &18.8      &56.9    &61.6        &121.0         &117.7\\[2ex]

Net\,Rate\,(s$^{-1}$)$^{\rm b}$  &$4.04_{-0.02}^{+0.02}$& $0.96_{-0.04}^{+0.04}$ &$0.156_{-0.002}^{+0.002}$&$0.053_{-0.001}^{+0.001}$&$0.051_{-0.001}^{+0.001}$\\ [2ex]

%\\
%Flux$_{(0.5-2)\rm keV}^{\rm c}$     
%			&4.69    &4.90      &4.33        &--             &--\\
%\\
%Flux$_{(2-10)\rm keV}^{\rm c}$      
%			&2.58	&2.71	  &--		   &3.46 		   &3.58\\
%\\
%Flux$_{(3-30)\rm keV}^{\rm c}$      
%			 &--		&--	      &--		   &4.60           &4.75\\
%\\
\hline
\end{tabular}
%\end{minipage}
\footnotesize  \caption{Summary of the 2016 simultaneous \xmmnu observation of \ton.}
\vspace{-5mm}
\begin{threeparttable}
\begin{tablenotes}
	\item[a] Net exposure time, after background screening and dead-time correction. Note that the RGS and MOS exposures are for both detectors combined. 
	\item[b] Net count rate between 0.4--10\,keV for EPIC-pn, 0.4--2\,keV for RGS in \xmm and 3--30\,keV for FPMA and FPMB in \nustar.
%	\item[c] Observed fluxes in the 0.5--2\,keV, 2--10\,keV and 3--30\,keV bands in units $\times10^{-12}$\,erg\,cm$^{-2}$\,s$^{-1}$.	
\end{tablenotes}
\end{threeparttable}
\label{tab:summary_obs}
\end{table*}
%\pagebreak
%\end{landscape} 

%\begin{landscape}
\begin{table*}
%\begin{minipage}{100mm}
\centering
%\scriptsize
\begin{tabular}{c@{\hspace{25pt}}c@{\hspace{15pt}}c@{\hspace{15pt}}c@{\hspace{15pt}}c@{\hspace{15pt}}c@{\hspace{15pt}}c}
\hline \\ [-2ex]

%Obs.~ID &  Instrument     &&&&&&\\

%                                                    &\multicolumn{3}{c}{\xmm}                                                                         &\multicolumn{2}{c}{\nustar}\\

Observation      &\multicolumn{2}{c@{\hspace{23pt}}}{$2000$}     &\multicolumn{2}{c@{\hspace{23pt}}}{$2002$}    &\multicolumn{2}{c@{\hspace{14pt}}}{$2015$}\\[2ex]

Telescope                          &\multicolumn{6}{c@{\hspace{15pt}}}{\xmm}                                                                                \\[2ex]

Obs.~ID &\multicolumn{2}{c@{\hspace{23pt}}}{$0110890401$} &\multicolumn{2}{c@{\hspace{23pt}}}{$0110890701$} &\multicolumn{2}{c@{\hspace{14pt}}}{$0764170101$}\\[2ex]

Instrument      &EPIC-pn               &RGS\,1$+$2       &EPIC-pn              &RGS\,1$+$2        &EPIC-pn             &RGS\,1$+$2 \\   [2ex]

Start Date     &2000--12--14     &2000--12--14     &2002--06--30    &2002--06--30   &2015--07--03     &2015--07--03\\[1ex]
 Time\,(UT)     &11:35:23	      &11:13:06         &03:06:41	     &03:00:57		    &22:32:33       &22:26:35 \\[2ex]

End Date       &2000--12--14     &2000--12--14     &2002--12--14    &2002--12--14   &2015--07--05     &2015--07--05\\[1ex]
Time\,(UT)      &19:45:22	     &19:46:29         &08:06:40		   &08:07:58		     &12:08:29       &12:05:20\\[2ex]

Duration\,(ks)            &29.4  &30.8             &18.0             &18.4              &135.5 		    &135.5 \\[2ex]

Exposure\,(ks)$^{\rm a}$  &18.9  &59.5             &11.9             &35.5              &84.0           &270.5  \\[2ex]

Net\,Rate\,(s$^{-1}$)$^{\rm b}$&$9.09_{-0.02}^{+0.02}$&$0.438_{-0.003}^{+0.003}$&$6.70_{-0.02}^{+0.02}$&$0.310_{-0.003}^{+0.003}$&$6.33_{-0.01}^{+0.01}$&$0.235_{-0.001}^{+0.001}$\\ [2ex]

%\\
%Flux$_{(0.5-2)\rm keV}^{\rm c}$     
%			&10.6            &10.3                  &7.79           &6.67               &7.12           &6.77\\
%\\
%Flux$_{(2-10)\rm keV}^{\rm c}$      
%			&5.06              &--                  &4.15           &--                 &3.95           &--\\
%\\
%Flux$_{(3-30)\rm keV}^{\rm c}$      
%			&--                &--                  &--             &--                 &--				&--\\
%\\
\hline
\end{tabular}
%\end{minipage}
\footnotesize  \caption{Summary of the 2000, 2002 and 2015 \xmm observations of \ton.}
\vspace{-5mm}
\begin{threeparttable}
\begin{tablenotes}
	\item[a] Net exposure time, after background screening and dead-time correction.
	\item[b] Net count rate between 0.4--10\,keV for EPIC-pn and 0.4--2\,keV for RGS.
%	\item[c] Observed fluxes in the 0.5--2\,keV and 2--10\,keV bands in units $\times10^{-12}$\,erg\,cm$^{-2}$\,s$^{-1}$.	
\end{tablenotes}
\end{threeparttable}
\label{tab:summary_obs_3epoch}
\end{table*}
%\pagebreak
%\end{landscape} 

%\pagebreak	

\section{Observations and Data Reduction}
\label{sec:Observations and Data Reduction}

\ton was observed between 10--13 June 2016 for a duration of $269\,\ks$ with \nustar, and then simultaneously with \xmm for $31\,\ks$ in the last part of the observation. For comparison, we also reduced the archival \xmm data of \ton from the observations carried out on 2000--12--14, 2002--06--30 and 2015--07--03,  with durations of 32, 18 and 141\,ks, respectively. 

\subsection{\xmm}

\subsubsection{EPIC}

 The \xmm EPIC instruments were operated in small window mode with the medium filter applied. The \xmm \citep{Jansen01} data were processed and cleaned by using the latest version of the Science Analysis System \citep[\sas v17.0.0; ][]{Gabriel04} and the most recent set of calibration files. As a standard procedure we filtered the EPIC data for background flares and the net exposure times are listed in Tables\,\ref{tab:summary_obs} and \ref{tab:summary_obs_3epoch}. The EPIC pn \citep{Struder01} and MOS \citep{Turner2001} source and background spectra were extracted from circular regions with radii of 40$\arcsec$ and two of 25$\arcsec$, respectively. 
The response matrices and ancillary files were generated with the \sas tasks \texttt{rmfgen} and \texttt{arfgen}. 
 
The source spectra were binned not to oversample the EPIC spectral resolution by a factor larger than 3, further imposing a minimum S/N of 5 per each energy bin.

\subsubsection{Reflection Gratings Spectrometer}
\label{subsub:RGS_reduction}

The \xmm Reflection Gratings Spectrometer data \citep[RGS hereafter;][]{Denherder2001} were reduced using the standard \sas task \rgsproc. The high-background intervals were filtered by applying a threshold of $0.2\,\rm cts\,s^{-1}$ on the background event files. We checked in each epoch that the RGS\,1 and RGS\,2 spectra were in agreement within the 3 per cent level. We then combined them by using the \sas task \rgscombine to a single RGS\,1$+$2 spectrum in each epoch (see Table\,\ref{tab:summary_obs_3epoch} for details).

Inspecting the EPIC and RGS spectra corresponding to the four epochs, there is no substantial change in the overall spectral shape but only a change in flux by a factor of $\sim$2.5 (see Section\,\ref{subsec:The broadband xmmnu analysis}). We then stacked all the four RGS\,1$+$2 spectra together into a single spectrum with $\sim10^5$ net counts. The stacked spectra were initially coarsely binned to a resolution of $\Delta\lambda = 0.1\ang$ per spectral bin. Such a binning is undersampling the RGS spectral resolution (i.e., $\Delta\lambda=0.06$--$0.08\,\ang$ full width at half maximum) over the $6$--$35\,\ang$ bandpass, however it can be useful to initially identify the presence of strong emission lines in the spectrum. Due to the high S/N ratio, we also binned the RGS spectra to $\Delta\lambda=0.03\ang$ for a more detailed analysis. 

\subsubsection{Optical Monitor}

In this work, we adopted the \xmm Optical Monitor telescope \citep[OM hereafter;][]{Mason01} in order to obtain the target's photometric points listed below. In the 2016 observation we obtained five $\sim$1 ks, 1.3 ks and 2.6 ks exposures, in imaging mode, through the V (effective wavelength $\lambda=5430\ang$), U ($\lambda=3440\ang$) and UVM2 ($\lambda=2310\ang$) filters, respectively. The data were processed by using the \sas \omi pipeline, which takes into account the all the necessary calibration processes such as flat-fielding. We also ran a source detection algorithm before performing aperture photometry on each detected source. The count rates were averaged over the different exposures and the up-to-date calibration uncertainties of the conversion factor between count rate and flux were added quadratically to the statistical error. 

Ideally, one should subtract the contribution of the host galaxy from the overall count rate. \ton largely resembles a point source where the AGN dominates over the galaxy emission. However, we do not have enough information on the host galaxy to exclude a non-negligible host contamination to the V band (see e.g., \citealt{Porquet19}). Only the U and UVM2 photometric points will be therefore used in this analysis.
Systematic errors of 1.4 and 1.5 per cent were respectively added to the U and UVM2 fluxes in quadrature\footnote{https://xmmweb.esac.esa.int/docs/documents/CAL-SRN-0346-1-0.pdf}. For each point we also applied an absorption correction due to the Galactic interstellar medium, based on the \citet{Cardelli89} extinction law. By using the standard value of $R_{V}=A_{V}/E(B-V)=3.1$ and setting $E(B-V)=0.0123$ from the Galactic extinction maps by \citet{Schlafly11}, we obtain $A_{V}=0.0383\pm0.0006\,\rm mag$, which translates into a count rate correction of $\sim1.06$ and $\sim1.11$ for the U and UVM2 filters, respectively. We used the OM canned response files\footnote{ftp://xmm.esac.esa.int/pub/ccf/constituents/extras/responses/OM} in order to analyse the OM photometry in \xspec.

\begin{figure}
  \includegraphics[width=0.45\textwidth]{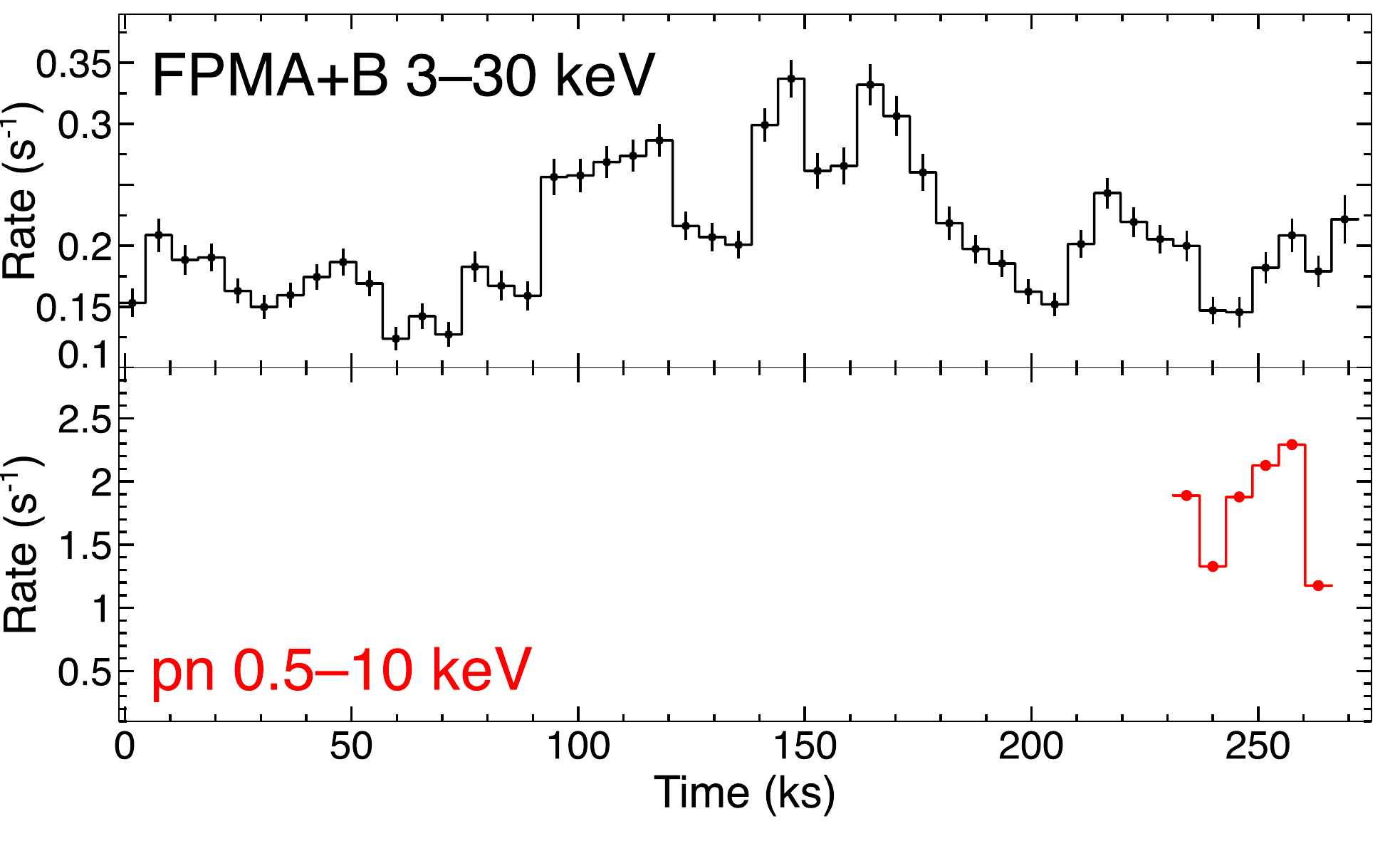}
  \caption{Top: Combined FPMA and FPMB background subtracted light curve in the $3$--$30\kev$ energy band showing the rapid variability during the $270\,\ks$ observation. Bottom: EPIC-pn light curve of \ton during the last $30\,\ks$ of the observation. For clarity the time bin of the light curves from both observatories correspond to the length of one \nustar orbit, i.e., $5814\,\rm s$.}
  \label{fig:ts180_2016_pn_nu_lc}
\end{figure}

\subsection{\nustar}

\ton was observed with \nustar \citep{Harrison13} between 10--13 June 2016 for a total duration of $\sim270\,\ks$, corresponding to a net exposure of about 120\,ks for both FPMA and FPMB detectors. Towards the end of the observation \ton was simultaneously targeted with \xmm, as shown in the respective light-curves plotted in \fig\ref{fig:ts180_2016_pn_nu_lc}.

The \nustar data were reduced according to the standard procedure by using the \heasoft task \nupip v0.4.6 of the \nustar Data Analysis Software package (\textsc{nustardas} v1.8.0). We used the most recent calibration files \textsc{caldb} v20180419 and then applied the standard screening criteria, such as the filtering for the South Atlantic Anomaly (SAA) by setting the mode \textit{optimized} in \nusaa v0.1.7. The spectra were extracted in each module from circular regions with radii of $40\arcsec$ and $90\arcsec$ for the source and background, respectively. For the spectral analysis in this paper, the resulting FMPA and FMPB spectra were rebinned so to oversample the intrinsic energy resolution (i.e., $\sim$400 eV over the range of interest) by a factor of 2.5, and further grouped to ensure a S/N of 5 per spectral channel. The light curves in the $3$--$30\kev$ energy band were extracted from the same regions using the \texttt{nuproducts} task. The background subtracted light curves from FPMA and FPMB were then combined into a single one.

\begin{figure}
  \includegraphics[width=0.45\textwidth]{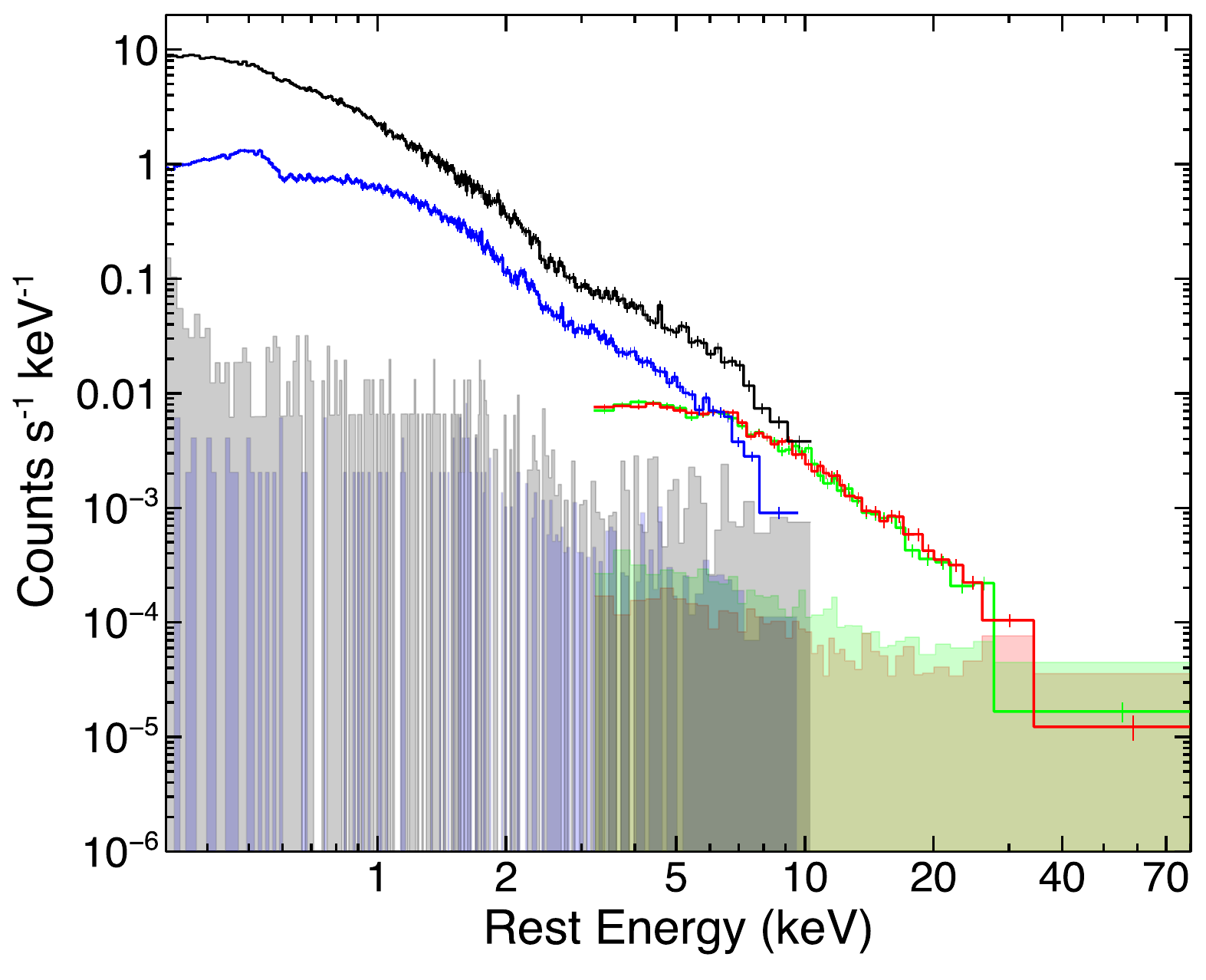}
  %\vspace{-50mm}
  \caption{Background subtracted EPIC-pn (black), MOS\,1$+$2 (blue), FPMA (red) and FPMB (green) spectra between $0.3$--$79\kev$ and the corresponding background levels of the joint 2016 \xmmnu observation of \ton. It is clear the the background contribution dominates above $\sim30\kev$. We will use the $0.4$--$30\kev$ energy band for the X-ray analysis in this paper.}
\label{fig:ts180_2016_ldata_bkg}
\end{figure}

\section{Spectral analysis}
\label{subsec:The broadband xmmnu analysis}

In \fig\ref{fig:ts180_2016_ldata_bkg} we show the background subtracted EPIC pn, MOS\,1$+$2 and FPMA/B spectra between $0.3$--$79\kev$, plotted with the corresponding X-ray background. The spectra have been corrected for the effective area of each detector. As \ton is characterized by a soft continuum, the hard band is background dominated at energies $E\gtrsim30$ keV. Regarding the soft X-ray band, the pn and MOS\,$1+2$ spectra diverge below 0.4 keV. In this work, we will then focus on the 0.4--30 keV range for the spectral analysis. In all fitting procedures, we include the contribution of a Galactic absorption column of $\nhgal=1.3\times10^{20}\,\rm cm^{-2}$, modelled with \texttt{Tbabs} \citep{Wilms00} and obtained from the recent H\,\textsc{i}\,$21\,\rm cm$ measurements \citep{HI4PI16}. We also assume solar abundance of the main elements throughout the analysis unless stated otherwise.

\begin{figure}
  \includegraphics[width=0.45\textwidth]{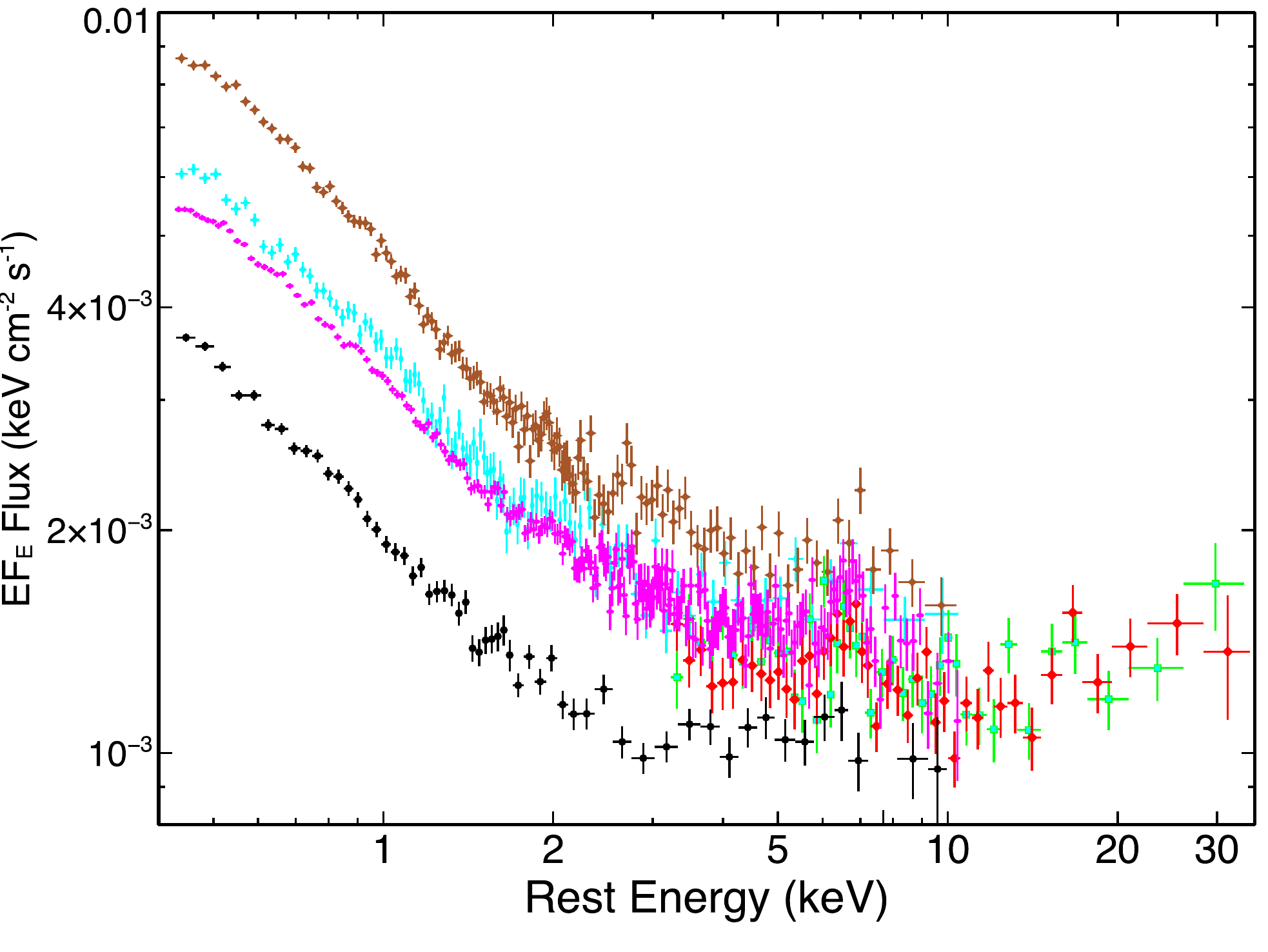}
  %\vspace{-50mm}
  \caption{Background subtracted \ton spectra corresponding to the 2016 \xmmnu EPIC-pn (black), FPMA (red) and FPMB (green). Note that the average intensity of the source over the entire \nustar observation was about 30 per cent larger than during the simultaneous coverage with \xmm (see Fig.\,\ref{fig:ts180_2016_pn_nu_lc}). Also the EPIC-pn spectra from 2000 (brown), 2002 (cyan) and 2015 (magenta) are shown for comparison.}
\label{fig:ts180_4b_ALL_eeuf}
\end{figure}

All the reduced spectra were analysed using the software packages \ftools v6.25 and \xspec v12.10.1b \citep{Arnaud96}. The spectra produced from all the detectors are either binned to a minimum significance of 5$\sigma$ (EPIC, FPMA/B) or characterized by $>$25 counts per channel (RGS), hence we adopted the $\chi^{2}$ minimization technique for the EPIC-pn, RGS and FPMA/B spectral analysis throughout this paper. While here we mostly focus on the 2016 \xmmnu spectra, for comparison purposes in \fig\ref{fig:ts180_4b_ALL_eeuf} also the \xmm (EPIC-pn) spectra of 2000, 2002 and 2015 are shown. It is evident that the 2016 \xmmnu observation (EPIC-pn and FPMA/B) was in the lowest flux state, but the observed variability is mostly driven by intensity changes, while the spectral shape below 10 keV is rather similar to previous epochs.

\begin{figure}
  \includegraphics[width=0.45\textwidth]{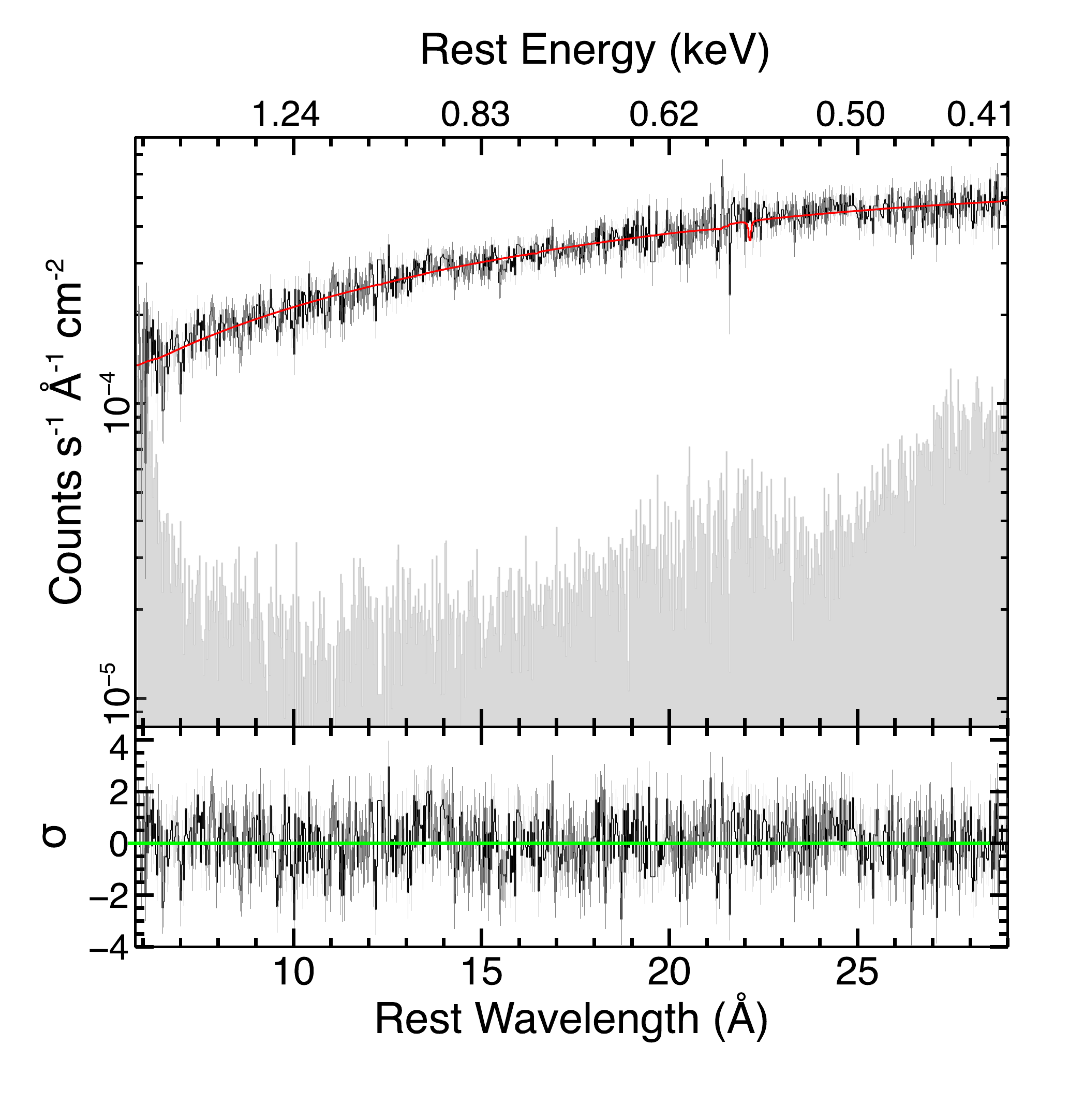}
  %\vspace{-50mm}
  \caption{Top: Stacked background subtracted RGS\,1$+$2 spectrum from the 2000, 2002, 2015 and 2016 \xmm observations of \ton. The best-fit (power-law modified by Galactic absorption) model is overlaid in red and the background level is shown in grey. Bottom: corresponding residuals of the data points compared to the best-fit model, in $\sigma$ units. The soft X-ray spectrum is largely consistent with a simple power law with a steep photon index of $\Gamma=2.93\pm0.02$.}
\label{fig:rgs_spectra}
\end{figure}

\begin{figure*}
  \includegraphics[width=0.75\textwidth]{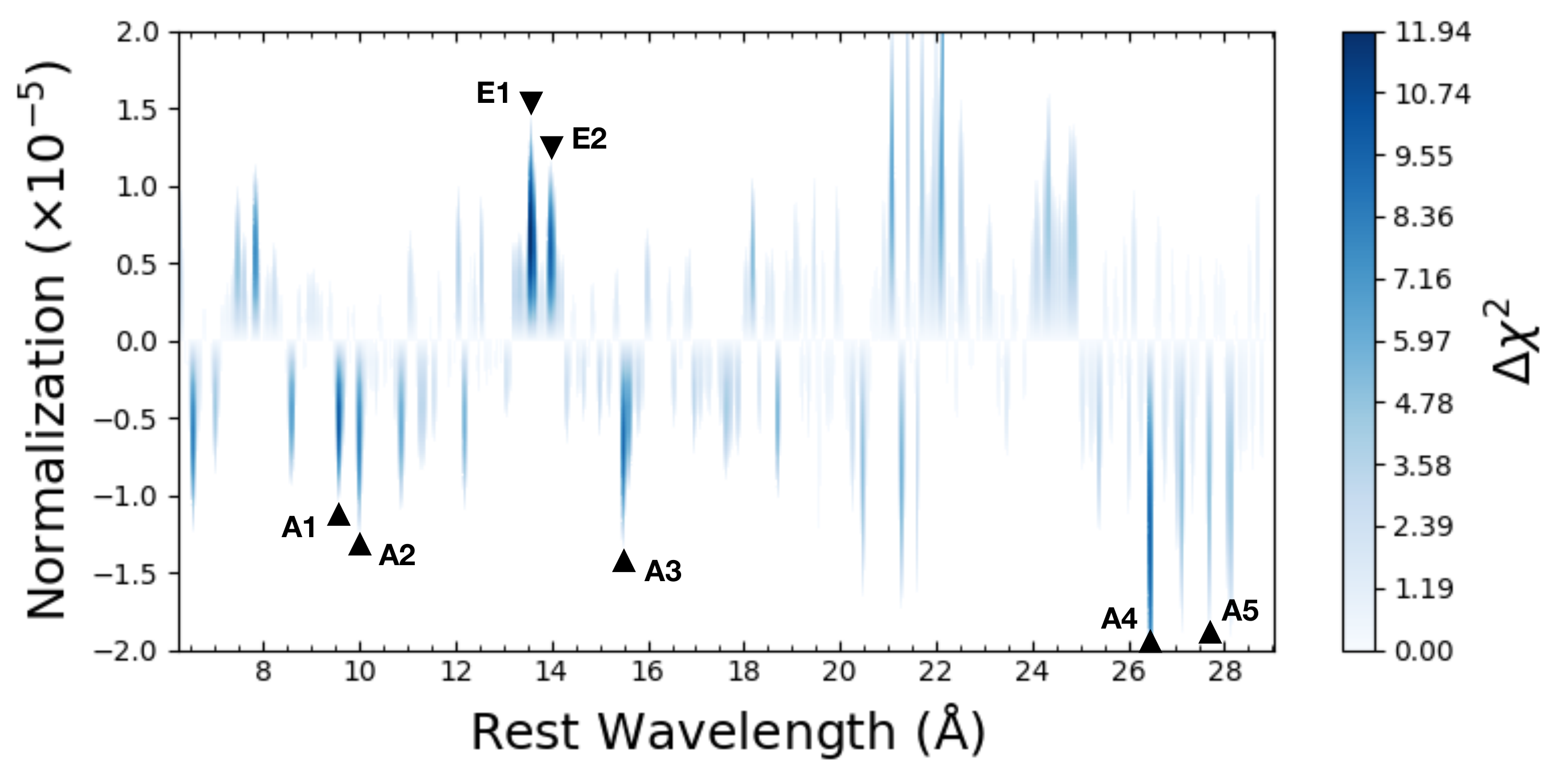}
  %\vspace{-50mm}
	\caption{Results of a blind line scan performed across the 6--31 $\ang$ band of the stacked RGS spectrum of \ton by adding a Gaussian profile in steps of 0.03 $\ang$ with free amplitude and width. The colour bar on the right indicates the maximum significance of the lines. A faint Ne\,\textsc{ix} triplet around 13.6 $\ang$ emerges in emission as the strongest feature, while the residuals in absorption, although individually not significant, suggest a marginal evidence of a wind with $\vout \sim -0.2c$. The most relevant emission and absorption features are flagged in the plot (see the text for more details).}
\label{fig:line_scan_rgs}
\end{figure*}

\begin{figure}
  \includegraphics[width=0.45\textwidth]{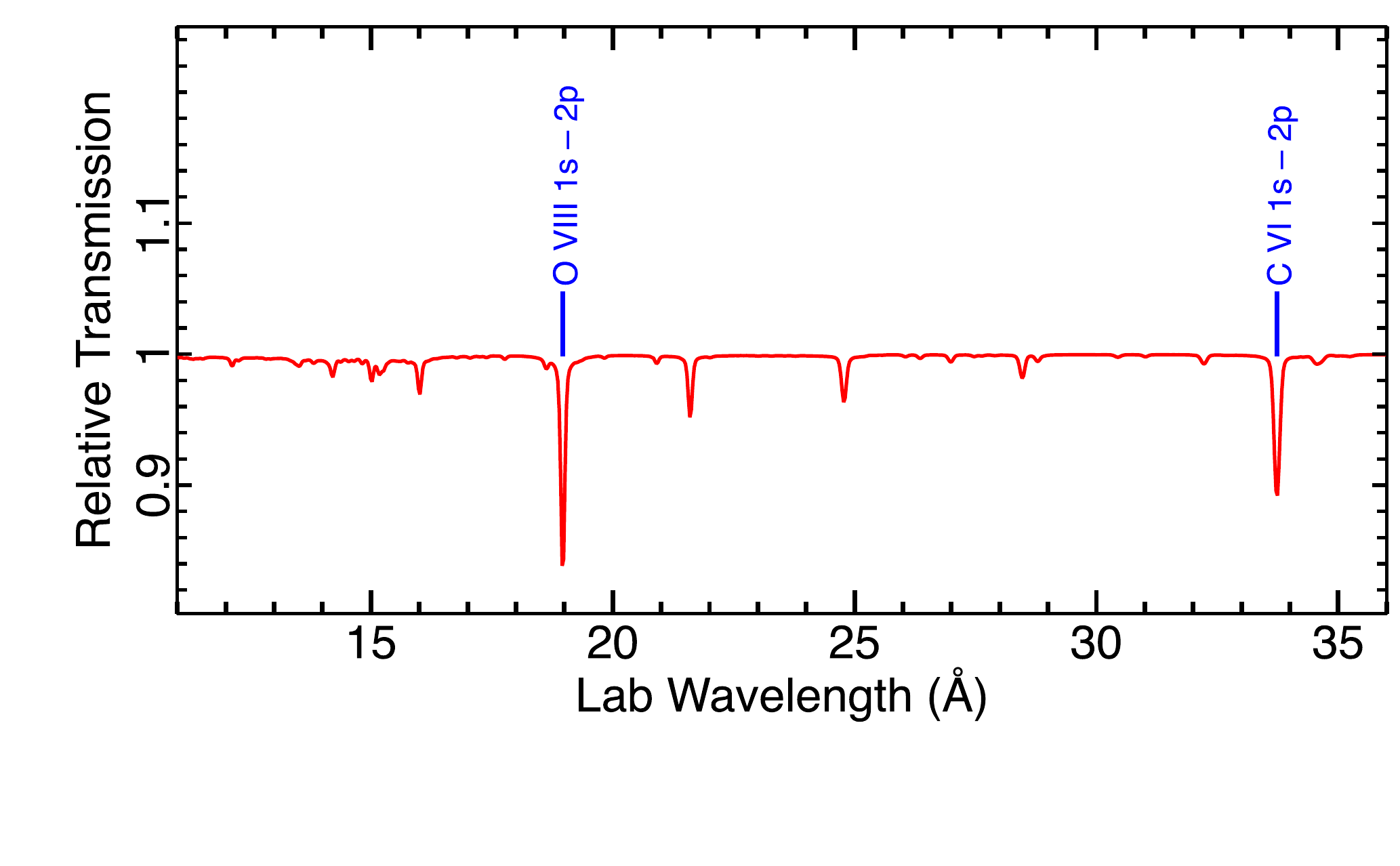}
  %\vspace{-50mm}
	\caption{Trasmission curve in the lab frame of the putative ionized absorber detected at the 3.1$\sigma$ confidence level in the RGS spectrum. The two main transitions are labelled, and both correspond to appreciable features in the line scan of Fig.\,\ref{fig:line_scan_rgs}, at $\lambda = 15.50$ (A3) and $27.71\,\ang$ (A5) in the rest frame of \ton. This would imply an outflow velocity of $\vout \sim -0.2c$.}
\label{fig:rgs_stack_xstar}
\end{figure}

\begin{figure}
  \includegraphics[width=0.45\textwidth]{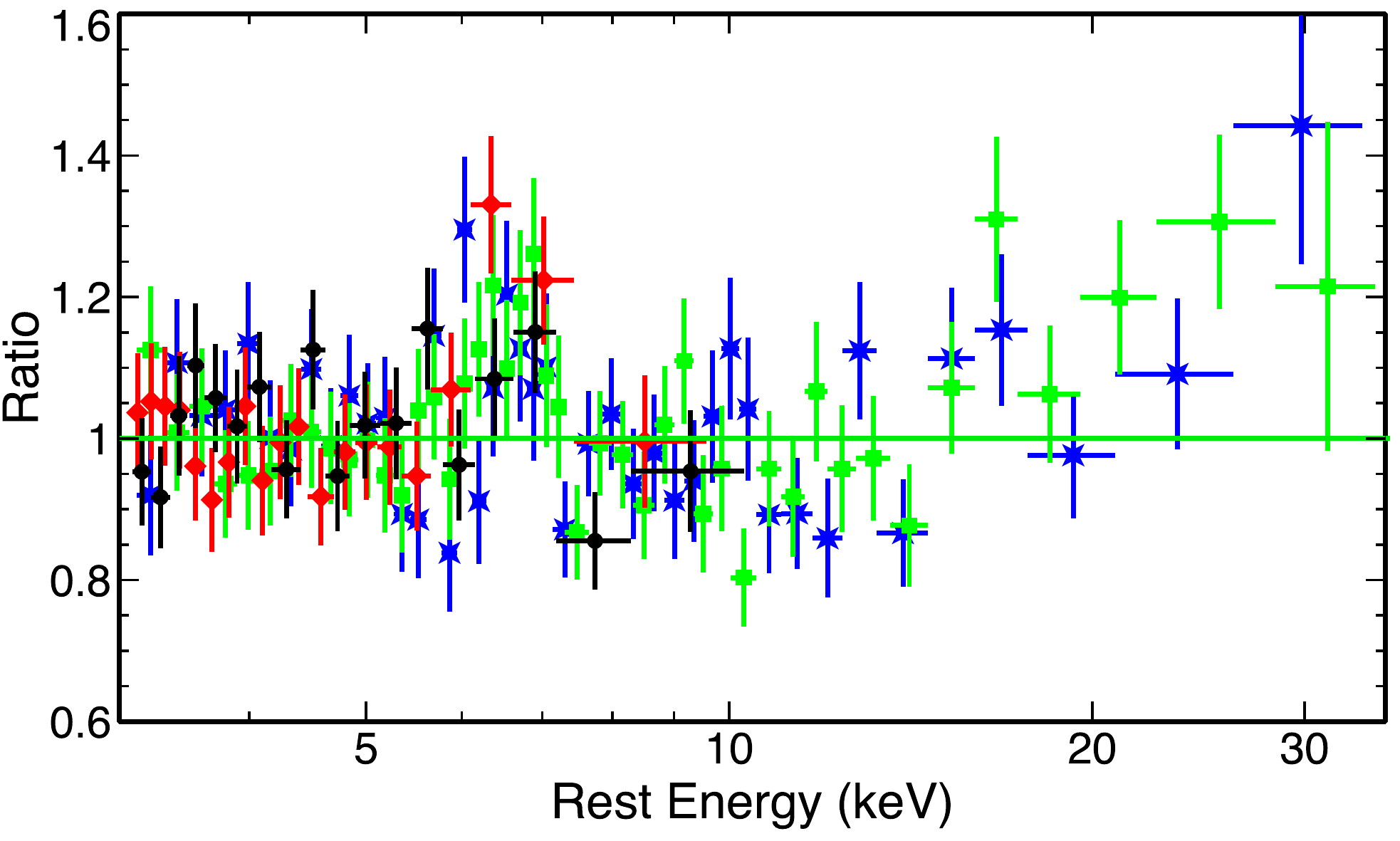}
  \caption{\xmmnu 2016 spectra in the $3$--$30\kev$ rest-energy band, fitted with a simple power law with photon index $\Gamma=2.07\pm0.03$. A broad, symmetric \fe emission component is detected at $\sim6.6$ keV, as well as some hard excess. Epic-pn, MOS, FPMA, FPMB are plotted in black, red, green and blue respectively.}
\label{fig:ts180_2016_3_30_pow}
\end{figure}

\subsection{RGS analysis}
\label{subsec:RGS analysis}

Previous analyses of \ton revealed that the soft X-ray excess below $\sim$2 keV is very steep and virtually featureless \citep[e.g.,][]{Turner01,Vaughan02b,Nardini12,Parker18ton}. When we inspected the 2016 RGS spectrum, which was caught in a low flux state, we did not detect any significant emission or absorption features. However, as illustrated in \fig\ref{fig:ts180_4b_ALL_eeuf}, the spectral shape of the soft excess in \ton hardly changes between the four \xmm observations in the sixteen years period. After a careful check of the individual RGS spectra, in order to increase the S/N, we combined them with \rgscombine into a single spectrum, which overall provides $1.105\pm0.004\times10^{5}$ net counts. This operation, in general, is not trivial (see, e.g., \citealt{Kaastra11b}). However, by simultaneously fitting the four individual spectra, we obtain qualitatively similar results.

At this stage we can therefore adopt a spectral binning of $\Delta\lambda=0.03\,\ang$, which fully samples the spectral resolution of the detector, and safely use the $\chi^{2}$ statistics. We find that modelling the spectrum with a simple power law and Galactic absorption produces an excellent fit, with $\chis=818/817$. In \fig\ref{fig:rgs_spectra} we show the stacked RGS\,1$+$2 spectrum with the power-law model overlaid in red. In the lower panel, the corresponding residuals in $\sigma$ units are shown. This result strongly suggests that the soft X-ray spectrum in \ton is largely consistent with a simple power-law with a steep photon index of $\Gamma=2.93\pm0.02$. Yet, given the data quality, we subsequently performed a blind emission and absorption line search by scanning the 6--31\,$\ang$ band in steps of 0.03\,$\ang$ for narrow Gaussian profiles. We included a line to the power-law continuum plus Galactic absorption model, allowing for both positive and negative normalizations and free (but limited by the resolution of the RGS spectrum) width. 

In \fig\ref{fig:line_scan_rgs} we show the results of our blind search, which confirms that the spectrum is largely featureless. The largest significance ($\dchidof=-11.9$/$-2$) is found at $\lambda_{\rm rest} = 13.55\,\ang$ (marked as E1 in \fig\ref{fig:line_scan_rgs}), which is formally the wavelength of the intercombination transition of the Ne\,\textsc{ix} triplet. Also the adjacent bins provide a significant improvement. Since we cannot distinguish the individual components of the triplet from the scan, we initially included in the baseline model a single Gaussian profile with free width. The fit improves by $\dchidof=-14.2$/$-3$ down to $\chis=804/814$ for $\lambda_{\rm rest} = 13.58\pm0.07\,\ang$ (consistent with the Ne\,\textsc{ix} intercombination component). Note that in this Section uncertainties are given at the 68 per cent ($\Delta \chi^{2}=1$) confidence level. The best line width and equivalent width are $\sigma \simeq 0.06^{+0.05}_{-0.04}\,\ang$ and ${\rm EW}=27\pm9$\,m$\ang$. We also tried to resolve the triplet by including three Gaussians with fixed $\sigma=0$ eV at the expected lab wavelengths of the resonant (r), intercombination (i), and forbidden (f) components. The statistical improvement in this case is slightly lower, $\dchidof=-11.2$/$-3$, for ${\rm EW(r)}<7$\,m$\ang$, ${\rm EW(i)}=17\pm6$\,m$\ang$, and ${\rm EW(f)}<9$\,m$\ang$. The only other barely significant emission feature in the line scan falls just shortwards of 14\,$\ang$ (labelled as E2). If real, this could be due to a blend of Fe-L transition, mostly from Fe\,\textsc{xxi}. The Ne\,\textsc{ix} detection supports the notion that even in (apparently) `bare' AGN, some gas with modest covering factor exists outside the line of sight \citep[see also][]{Reeves16b}.

Some caution is required in interpreting the possible absorption features in the stacked spectrum, which could be found at unsual energies if associated with outflowing material. We first repeated the line scans on the single epochs, verifying that the 2015 observation dominates the statistics. Three lines in the stack are significant at more than the 99 per cent level ($\dchidof<-9.2$/$-2$) according to the blind scan, at $\lambda_{\rm obs} = 9.57$ (A1), $15.50$ (A3), and $26.46\,\ang$ (A4). None of these correspond to strong transitions in either the rest frame of \ton or the lab frame. When the number of resolution elements is taken into account (a spurious $\Delta\chi^2$ improvement can appear in any bin), no safe detection is confirmed. Indeed, \citet{Parker18ton} did not find any significant features in the 2015 RGS spectrum. However, in order to constrain the properties of any possible ionized absorber along the line of sight, we adopted the \xstar photoionization code \citep{BautistaKallman01,Kallman04} to generate an absorption table with a realistic ionizing SED (see Section\,\ref{subsec: two-corona broadband modelling with AGNSLIM}) as input spectrum, gas density of $n=10^{12}$ cm$^{-3}$ (from the intensity ratio between forbidden and intercombination lines in the Ne\,\textsc{ix} triplet, after \citealt{PorquetDubau00})\footnote{This estimate should be taken with caution, since the constraints on the Ne\,\textsc{ix} components are rather poor, and the assumption that emission and absorption arise from the same gas is not necessarily true. The results, however, are only mildly sensitive to the choice of the gas density.}, and
turbulent broadening of 300 $\kms$ (based on the resolution of the RGS spectrum). The \xstar code computes the radiative transfer through a spherically symmetric shell of ionized gas. Through the generated absorption grid we can measure three physical parameters of the gas: the column density $\nh$, the ionization parameter\footnote{This is defined as $\xi=\lion/nR^{2}$ \citep{Tarter69}, where $\lion$ is the ionizing luminosity, $n$ is the gas number density and $R$ is the radial distance from the ionizing source.}, and the outflow velocity (via the redshift parameter). 

The inclusion of the \xstar grid improves the fit by $\dchidof=-14.7$/$-3$, equivalent to a 3.1$\sigma$ confidence. The column density is largely unconstrained between $\log(\nh/\rm cm^{-2})=$19.7--22.0, while $\logxi = 2.7_{-0.1}^{+0.2}$ and $\vout \simeq -0.195c$.
Assuming that the absorber is instead local to our Galaxy or to the AGN frame gives a worse fit by $\Delta\chi^2=9$ and 15, respectively. 

In Fig.\,\ref{fig:rgs_stack_xstar} we show the transmission curve of the putative absorber shifted to the lab frame. Interestingly, the two main lines from O\,\textsc{viii} (1s$\to$2p, $\lambda_{\rm rest} = 18.97\,\ang$) and C\,\textsc{vi} (1s$\to$2p, $\lambda_{\rm rest} = 33.74\,\ang$) 
correspond to major features in the line scan at $\lambda_{\rm obs} = 15.50$ and $27.71\,\ang$ (A5). At this outflow velocity, also the $\lambda_{\rm obs} = 9.57$, $10.02$ (A2) and $26.46\,\ang$ residuals can be associated with resonant transitions, from Fe\,\textsc{xxii} (2p$\to$3d, $\lambda_{\rm rest} = 11.77\,\ang)$, Fe\,\textsc{xxi} (2p$\to$3d, $\lambda_{\rm rest} = 12.28\,\ang$) and Li-like S\,\textsc{xiv} (2p$\to$3d, $\lambda_{\rm rest} = 32.38\,\ang)$, but these cannot be reproduced by the same absorption grid as they require different ionization (that is, higher in the case of the Fe-L complex).

The presence of a fast accretion disc wind in \ton, although intriguing, remains tentative (see also Section\,\ref{sec:wind}). In any case, we can consider the soft X-ray spectrum as effectively featureless in the following analysis of CCD-resolution spectra.

\begin{figure*}
  \includegraphics[width=0.75\textwidth]{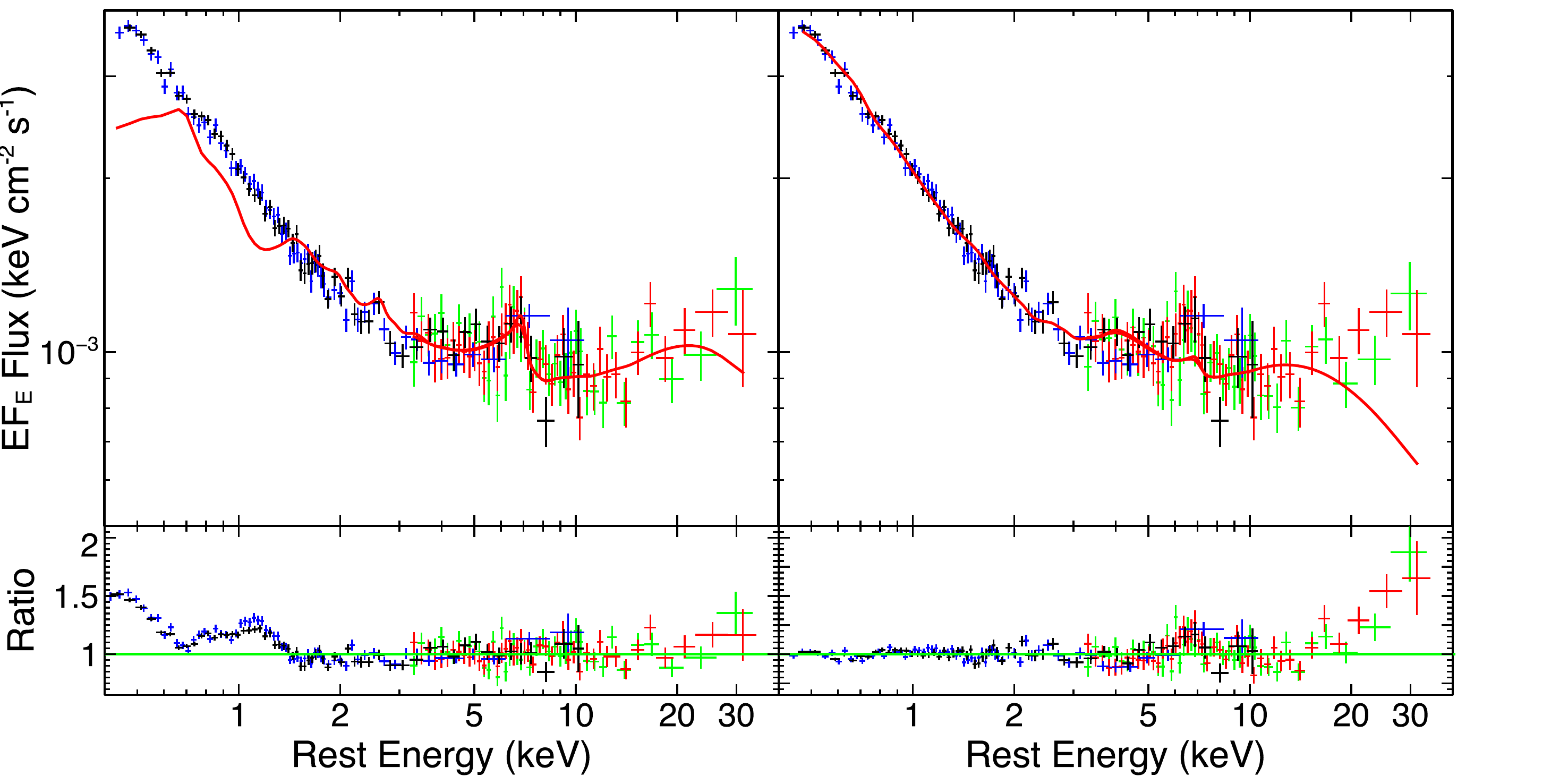}
  %\vspace{-50mm}
  \caption{The 2016 \xmmnu spectra of \ton fitted with \relxill where the overall model (solid red line) is overlaid to the data. Left: fit in the $3$--$30$ keV energy range where the \fe profile and the hard X-ray data are well reproduced. However, the extrapolation down to 0.4 keV of the model fails to account for the soft excess leaving strong residuals below 2 keV. Right: fit in the 0.4--30 keV energy band. While the soft excess is now well accounted for, the model cannot adequately fit the \fe emission nor the hard X-rays at $E > 10$ keV. Fitting the \xmmnu spectra with a standard reflection model cannot account self-consistently for the soft and hard X-ray bands. The values for both cases are tabulated in Table\,\ref{tab:broadband_relxill_no_xillver_2016}.}
\label{fig:ts180_2016_relxill_only}
\end{figure*}

\begin{figure}
  \includegraphics[width=0.45\textwidth]{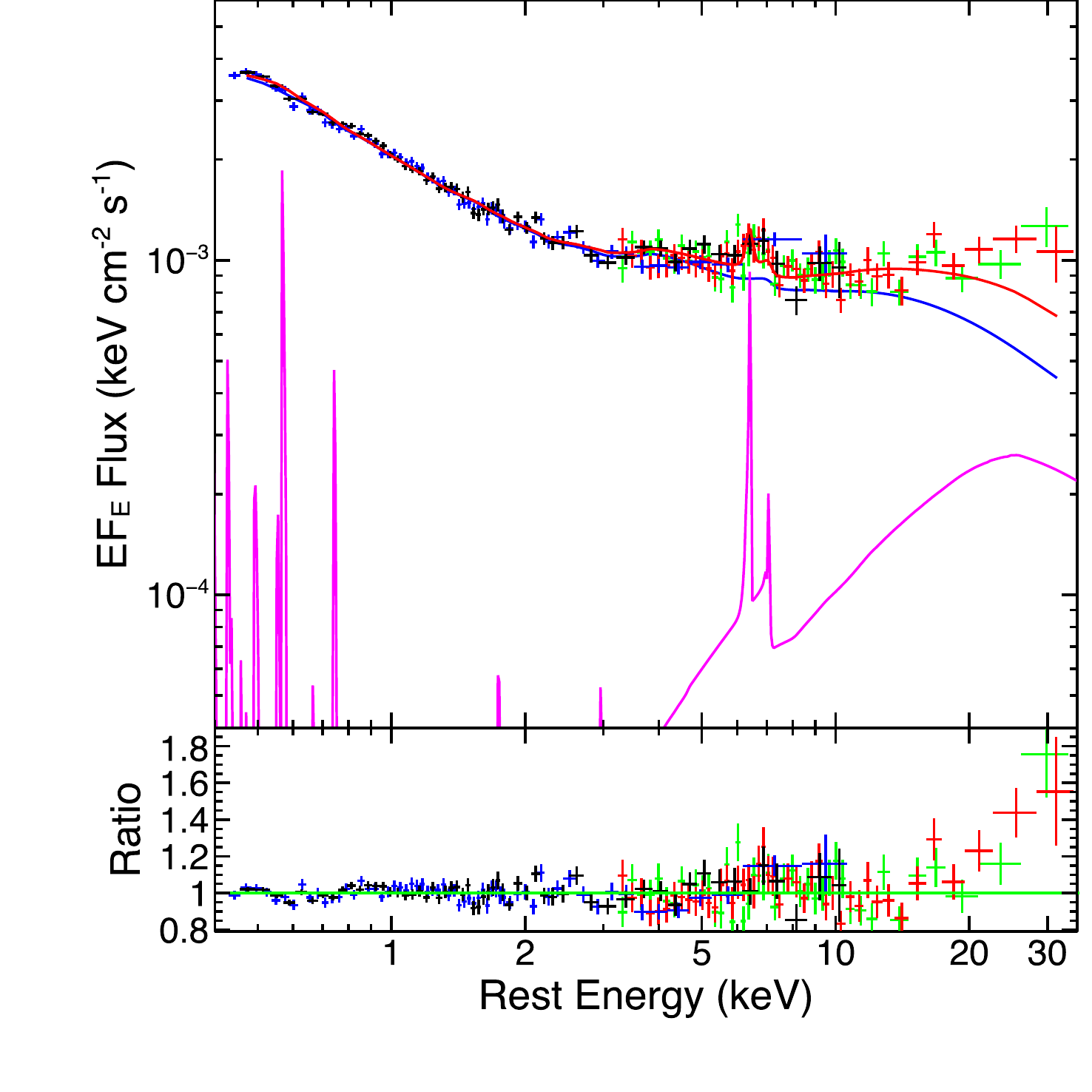}
  \vspace{-5mm}
  \caption{Same as in \fig\ref{fig:ts180_2016_relxill_only}\,(right panel), but with the addition of a distant reflector component modelled with \xillver (magenta) to the blurred reflection (blue), here reproduced by \relxillcp, in order to reduce the residuals at $E > 10$ keV. Despite the considerable statistical improvement, the best-fit model still struggles to properly account for the \fe and hard X-ray band (see Table\,\ref{tab:broadband_relxill_models_2016} for the model parameters).}
\label{fig:ts180_relx_eeuf_ra}
\end{figure}

\begin{figure}
  \includegraphics[width=0.45\textwidth]{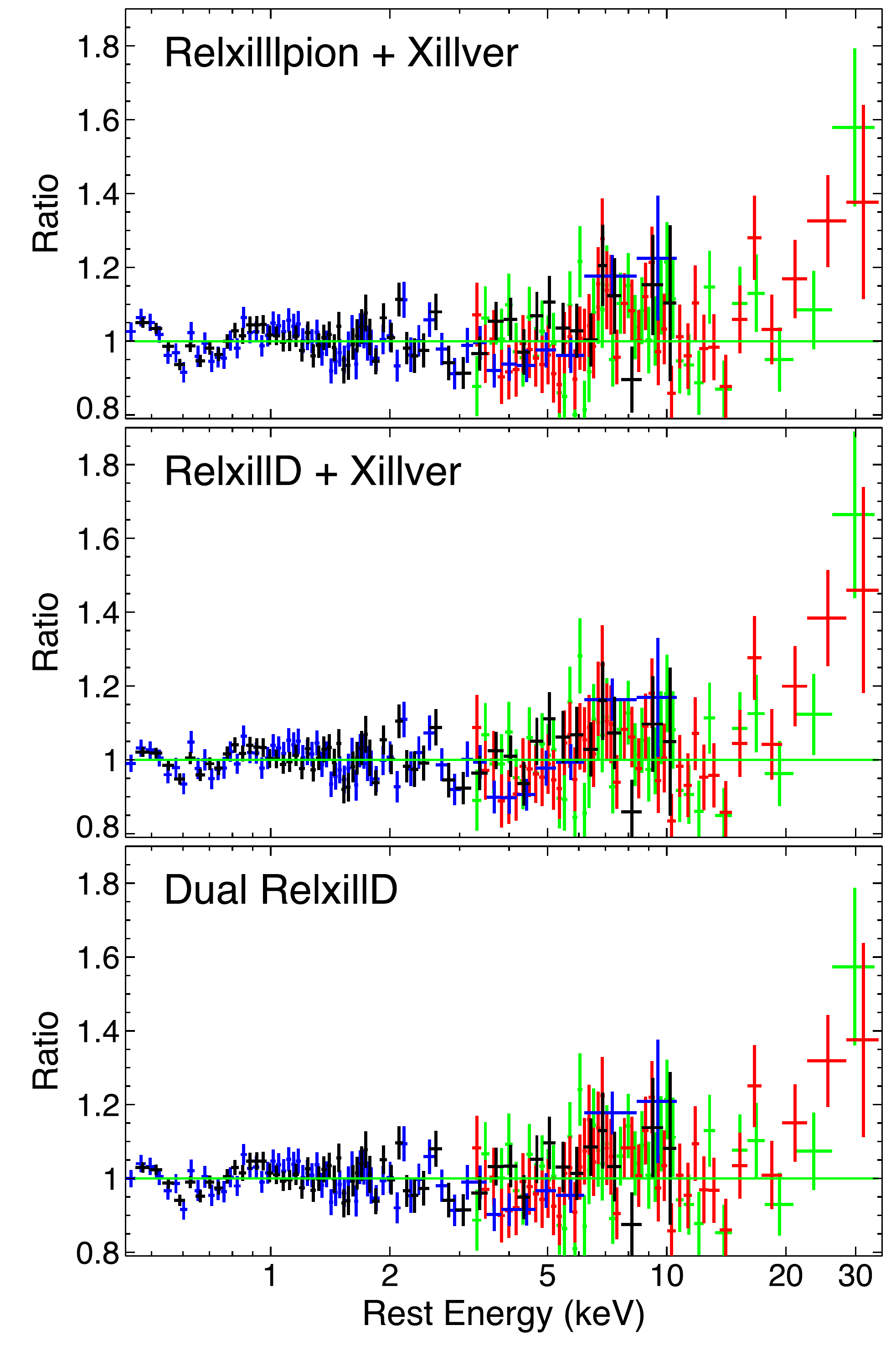}
%  \vspace{-20mm}
  \caption{Data/model ratio plots corresponding to three different fits carried out on the 2016 \xmmnu data of \ton with different flavours of the \relxill relativistic reflection models. Relativistic reflection alone, regardless of which version of \relxill we use, cannot self-consistently account for the $0.4$--$30\kev$ spectra leading to significant residuals in the \iron band and above $10\kev$.}
\label{fig:ts180_2016_relLP_CP_D_ra}
\end{figure}

%%%%%%%%%%%%%%%%%%%%%%%%%%%%%%%%% TABLE WITH 2016 RELXILL ONLY 3-30 0.4-30 MODELS %%%%%%%%%%%%%%%%%

\begin{table*}

\begin{tabular}{c@{\hspace{15pt}}c@{\hspace{25pt}}c@{\hspace{25pt}}c@{\hspace{25pt}}c}

\hline

Component                   &Parameter             &$3$--$30\kev$ &$0.4$--$30\kev$   &Description\,(units) \\

\hline \\ [-2ex]

%    \multirow{1}{*}{\texttt{Tbabs}}

\texttt{Tbabs}				     &$\nhgal$    &\multicolumn{2}{@{\hspace{-20pt}}c}{$1.3\times10^{20}$}&Galactic column\,(\cmsq)\\

\\

%\multirow{11}{*}{}

\relxill				    &$q_1$          &$3^{*}$   &$>9.3$                 &Inner emissivity index\\[1ex]

                                    &$q_2$          &$-$     &$3^{*}$                &Outer emissivity index\\[1ex]

					                &$R_\rmn{br}$            &$-$  &$3.7_{-0.2}^{+0.6}$    &Break radius\,($\rg$)\\[1ex]

					                &$a$            &$<0.72$ &$0.94_{-0.05}^{+0.03}$  &Black hole spin\\[1ex]

					                &$i$           &$41.6_{-3.7}^{+5.2}$    &$40.1_{-8.7}^{+5.5}$   & Inclination\,(degrees)\\[1ex]

					                &$\Gamma$      &$2.31_{-0.07}^{+0.07}$  &$2.37_{-0.01}^{+0.04}$  &Photon index \\[1ex]

				                   &$E_{\rm cut}$  &$300^{*}$                   &$300^{*}$                  &High energy cut-off\,(keV)\\[1ex]

					        &$\log(\xi)$    &$2.3^{+0.3}_{-0.5}$      &$2.9_{-0.3}^{+0.1}$     &Ionization\,(erg\,cm\,s$^{-1}$)\\[1ex]

				                   &$A_{\rm Fe}$   &$1.0^{*}$      &$0.9_{-0.1}^{+1.8}$     &Iron abundance\,(solar) \\[1ex]

				                   &$\mathcal{R}$  &$0.57^{+0.27}_{-0.19}$      &$2.0_{-0.4}^{+0.9}$                 &Reflection fraction\\[1ex]				
					                      
	   &norm  &$1.7_{-0.2}^{+0.2}\times10^{-5}$ &$1.0_{-0.1}^{+0.6}\times10^{-5}$    &Normalization (arbitrary) \\					                     	                      					                     

\\

                                  &MOS   &$1.03\pm0.04$          &$1.05\pm0.01$      &\multirow{3}{*}{Cross-normalization}\\

                                  &FPMA  &$1.28\pm0.05$          &$1.29\pm0.04$     \\

                                  &FPMB  &$1.32\pm0.05$          &$1.34\pm0.04$     \\

\hline

Fit statistic            &$\chi^2/\nu$   &$275.2/278$           &$522.8/410$\\

\hline

\end{tabular}

\caption{Summary of the best-fit parameters of the standard relativistic reflection model applied for the 2016 \xmmnu spectra of \ton over the 3--30 keV and 0.4--30 keV energy range (see text for details). $^{*}$ denotes a frozen parameter during fitting.}
\vspace{-5mm}
\label{tab:broadband_relxill_no_xillver_2016}
\end{table*}

%%%%%%%%%%%%%%%%%%%%%%%%%%%%%%%%% TABLE WITH 2016 RELXILL BROADBAND MODELS %%%%%%%%%%%%%%%%%

\begin{table*}

\begin{tabular}{c@{\hspace{10pt}}c@{\hspace{15pt}}c@{\hspace{15pt}}c@{\hspace{15pt}}c@{\hspace{15pt}}c@{\hspace{15pt}}c}

\hline

	Component                        &Parameter             &--\texttt{Cp} &--\texttt{lpion} &--\texttt{D} &--\texttt{D} ($\times2$)  &Description\,(units)\\

\hline \\ [-2ex]

%    \multirow{1}{*}{\texttt{Tbabs}}

\texttt{Tbabs}				     &$\nhgal$     &\multicolumn{4}{c@{\hspace{20pt}}}{$1.3\times10^{20}$}&Galactic column\,(\cmsq)\\

\\

%\multirow{14}{*}{\texttt{relxill}--}

\texttt{relxill}--                                    &$q_1$          &$>9.1$                 &--    &$>9.1$                &$>9.1$&Inner emissivity index\\[1ex]

                                    &$q_2$          &$3^{*}$                &--    &$3^{*}$               &$2.0_{-0.4}^{+3.7}$&Outer emissivity index\\[1ex]

					                &$R_\rmn{br}$            &$3.7_{-0.3}^{+0.5}$    &--    &$3.7_{-0.3}^{+0.3}$   &$4.8_{-2.0}^{+27.5}$&Break radius\,($\rg$)\\[1ex]

                                    &$h$           &--                      &$<2.1$     &--&--&Source height\,($\rg$)\\[1ex]

					                &$a$            &$0.95_{-0.04}^{+0.02}$ &$>0.98$  &$0.95_{-0.02}^{+0.01}$    &$>0.98$&Black hole spin\\[1ex]

					                &$i$           &$40.3_{-3.5}^{+6.3}$    &$28.7_{-4.8}^{+4.1}$   &$40.2_{-7.1}^{+5.6}$     &$57.8_{-6.4}^{+2.2}$& Inclination\,(degrees)\\[1ex]

					                &$\Gamma$      &$2.40_{-0.02}^{+0.02}$  &$2.44_{-0.01}^{+0.02}$  &$2.42_{-0.03}^{+0.02}$     &$2.48_{-0.02}^{+0.05}$&Photon index \\[1ex]

											                   &$E_{\rm cut}$  & $-$                   &$300^{*}$                  & $300^{\rm df}$&$300^{\rm df}$  &High energy cut-off\,(keV)\\[1ex]

					               &$kT_{\rm e}$   & $100^*$                      &--                      &--        &--  &Coronal temperature\,(keV)\\[1ex]

					        &$\log(\xi)$    &$3.0_{-0.1}^{+0.1}$      &$3.0_{-0.2}^{+0.1}$     &$3.0_{-0.1}^{+0.1}$     &$3.0_{-0.1}^{+0.1}$&Ionization\,(erg\,cm\,s$^{-1}$)\\[1ex]
					                     					    						   
&$p_{\xi}$   &$-$      & $<0.03$     &$-$     &$-$ & Ionization gradient index \\[1ex]

				                   &$A_{\rm Fe}$   &$0.9_{-0.1}^{+0.1}$      &$2.3_{-0.6}^{+0.7}$     &$0.9_{-0.1}^{+0.3}$     &$2.4_{-0.6}^{+0.2}$&Iron abundance\,(solar) \\[1ex]			
						       &$\log(n_1)$   &$15^{\rm df}$    &$15^{\rm df}$    & $<16.0$                    &  $<15.1$ &(Inner) disc density\,(\cmq)\\[1ex]	
				                     
&$\log(n_2)$   &--    &--    & --                    & $16.4^{+0.2}_{-0.3}$ &Outer disc density\,(\cmq)\\[1ex]

				                   &$\mathcal{R}$  &$2.1_{-0.5}^{+2.4}$      &$-$                 &$1.5_{-0.4}^{+4.1}$     &$1.3_{-0.1}^{+2.3}$&Reflection fraction\\[1ex]				

%					               &Fix$\mathcal{R}$  &--                    &$1^{*}$                 &--                      &--&Fix Reflection fraction\\

	   &norm$_1$  &$0.9_{-0.4}^{+0.2}\times10^{-5}$ &$5.3_{-0.8}^{+0.8}\times10^{-4}$    &$1.3_{-0.3}^{+0.2}\times10^{-5}$     &$1.1_{-0.2}^{+0.1}\times10^{-5}$&Normalization (arbitrary)\\[1ex]

&norm$_2$  &-- &--    &--     &$0.7_{-0.2}^{+0.3}\times10^{-5}$&Normalization (arbitrary)\\

\\

%\multirow{1}{*}{\xillver}

\xillver	&norm  &$1.7_{-0.5}^{+0.6}\times10^{-5}$   &$1.6_{-0.3}^{+0.3}\times10^{-5}$    &$1.6_{-0.5}^{+0.4}\times10^{-5}$  &$1.8_{-0.5}^{+0.5}\times10^{-5}$&Normalization (arbitrary)\\

\\

                                  &MOS   &$1.05\pm0.01$          &$1.05\pm0.01$      &$1.05\pm0.01$      &$1.05\pm0.01$&\multirow{3}{*}{Cross-normalization}\\

                                  &FPMA  &$1.29\pm0.04$          &$1.32\pm0.04$      &$1.30\pm0.04$      &$1.30\pm0.04$\\

                                  &FPMB  &$1.34\pm0.04$          &$1.37\pm0.04$      &$1.34\pm0.04$      &$1.34\pm0.04$\\

\hline

Fit statistic            &$\chi^2/\nu$   &$496.6/409$           &$541.6/410$        &$491.0/408$      &$492.1/405$\\

\hline

\end{tabular}

\caption{Summary of the broadband best-fit parameters obtained by applying the advanced configurations of the \relxill family of reflection models to the 2016 \xmmnu spectra of \ton (see text for details). $^{*}$ and $^{\rm df}$ respectively denote a frozen parameter during fitting and a fixed setting hardwired into a given model component. For the \xillver component, $A_{\rm Fe}=1$, $\logxi=0$, and $E_{\rm cut}=300$ keV are assumed.}
\vspace{-5mm}
\label{tab:broadband_relxill_models_2016}
\end{table*}
%\pagebreak

\subsection{Broadband modelling with relativistic reflection}
\label{subsec:Modelling with relativistic reflection}

Given the lack of strong emission (or absorption) features in the RGS spectrum, we now investigate whether X-ray reflection or Comptonization models can account for the broadband spectrum of \ton. Reflection models have been successful in reproducing self-consistently the prominent features imprinted on the X-ray spectrum of AGN and X-ray binaries. These include the soft and hard X-ray excesses as well as broad \fe emission profiles \citep[e.g.,][]{Fabian04,Fabian10,Walton13,Jiang19}. In \fig\ref{fig:ts180_2016_3_30_pow} we show the 2016 spectra in the $3$--$30\kev$ rest-energy band, fitted with a simple power law with photon index $\Gamma=2.07\pm0.03$. A broad ($\sigma\sim300\ev$) and symmetric \fe emission component (with centroid energy at $E_{\rm rest}\sim6.6$ keV) is evident, as well as some hard excess above $\sim15\kev$. As a first step, we aim to test whether these features can be reproduced with a relativistic reflection model alone. 

We start by fitting the $3$--$30\kev$ band, where a cross-normalization between the EPIC and FPMA/B spectra is included to account also for the slightly different average flux. We adopted the state-of-the-art reflection model \relxill v1.2.0 \citep{Garcia14,Dauser14}, with Fe abundance fixed to solar and emissivity index $q=3$, while spin, disc inclination, ionization parameter and reflection fraction were allowed to vary. The disc inner radius is set by the spin parameter, i.e., $R_{\rm in,disc} \equiv R_{\rm isco}$.\footnote{Later in this work, we will distinguish between the `physical' disc inner radius $R_{\rm in,disc}$ and a `reflection inner radius', $R_{\rm in,refl}$. The two quantities are coincident in pure reflection models.} We find that the model reproduces very well the $3$--$30\kev$ spectra with $\chis=275/278$. The best-fit parameters are listed in Table\,\ref{tab:broadband_relxill_no_xillver_2016}. However, by extrapolating the \xmm spectra down to 0.4 keV the basic \relxill model is not able to account for the soft X-ray excess, as shown in \fig\ref{fig:ts180_2016_relxill_only}\,(left). Given that the \nustar spectrum is truncated at 30 keV due to the high background, we cannot constrain the cut-off energy of the primary continuum (hence the hot-corona temperature), which was subsequently fixed to 300 keV for the remainder of the paper.

\vspace{1cm}

The broadband ($0.4$--$30$ keV) spectrum was then refitted allowing for a broken power-law emissivity function with break radius $R_\rmn{br}$ and for free Fe abundance. In the fitting procedure, we fixed the outer emissivity index to the classical (non relativistic) limit of $q_{2}=3$, and keep this assumption for the rest of the reflection analysis, unless stated otherwise. This broken power-law configuration resulted in a break radius of $R_{\rm br}\sim4\,\rg$, which could be considered as a stringent lower limit for the size of the corona, considering a reasonable extension of $\sim$10\,$\rg$ \citep[e.g.,][see also Section\,\ref{subsec:Broadband modelling with optxagn}]{Wilkins12,Kammoun19}. Statistically, the new model returned a fairly decent fit, $\chis=523/410$. However, whilst fitting well the soft excess, this model is incapable of accounting for the \fe emission profile and the hard X-rays above $\sim$10 keV. This is because, after the inclusion of the soft band, more extreme parameters (e.g., $q_{1}\gtrsim9$, $a\sim0.94$; Table\,\ref{tab:broadband_relxill_no_xillver_2016}) are needed to reproduce the smoothness (see Section\,\ref{subsec:RGS analysis}) and steepness of the soft X-ray spectrum. Indeed, also the slope of the primary continuum slightly increases, so that significant residuals are present over the $3$--$30$ keV band (where now $\chis=358/275$), as illustrated in \fig\ref{fig:ts180_2016_relxill_only}\,(right). 

To determine whether reflection can explain the 2016 \xmmnu data, we must therefore allow for more complex models, involving the different physical and geometrical configurations of the disc/corona available within the \relxill package. Firstly, we attempt to improve the above fit by including the additional contribution of a neutral, unblurred reflector, modelled with \xillver \citep{Garcia13}. Previous works established that evidence of a distant reflection component (i.e., from the torus) in \ton is at most marginal \citep[e.g.,][]{Takahashi10,Nardini12,Parker18ton}, as a narrow \fe core is virtually undetected. Nonetheless, the high S/N data provided by the \nustar coverage allows us to better investigate the true level of the continuum above 10 keV. Secondly, for the disc/corona component we switch to the \relxillcp version, in which the reflection spectrum is calculated by using a more physical primary continuum, implemented with the \texttt{nthcomp} Comptonization model \citep{Zdziarski96,Zycky99} instead of a simple cut-off power law. The `seed' photon temperature is fixed at 50 eV. The  model has virtually the same parameters as \relxill, with the only difference that in the former the high energy cut-off $E_{\rm cut}$ is replaced by the coronal electron temperature, $kT_{\rm e}$. An approximated relationship between these two quantities is $E\simeq3kT_{\rm e}$ \citep{Petrucci01}. Accordingly, we can directly associate the cut-off energy to the electron temperature. 

The overall relativistic plus distant reflection model can be described as \texttt{Tbabs}*(\relxillcp$+$\xillver).\footnote{The disc inclinations and the photon indices are tied together between the `unblurred' and `blurred' reflector components. Solar iron abundance and $\logxi=0$ are assumed for the former component.} The addition of the unblurred component provided a considerable improvement to the fit, by $\dchidof=-26$/$-1$ (i.e., $\sim$5$\sigma$), leading to $\chis=497/409$. Despite such improvement, it appears that the model still struggles to properly account for both the \fe region (between $\sim$6--8 keV) and the hard X-rays above 10 keV, as shown in \fig\ref{fig:ts180_relx_eeuf_ra}.  The \relxillcp$+$\xillver model yielded a reflection fraction of $\mathcal{R}\sim2$, a disc inclination of $i=40_{-3}^{+7}\,\rm deg$, and a disc ionization of $\logxi=3.0\pm0.1$, which are all broadly consistent with what \citet{Parker18ton} found for the 2015 \xmm observation. Iron abundance is consistent with the solar value, $A_{\rm Fe}=0.9\pm0.1$. In terms of the primary continuum parameters, this model returned a steep photon index of $\Gamma=2.40\pm0.02$, which is consistent with what was measured in previous observations of \ton \citep[e.g.,][]{Comastri98,Turner98,Nardini12,Parker18ton}. Note that this value is steeper by $\Delta\Gamma \sim 0.1$ than what is derived from the $3$--$30\kev$ band. This confirms that the soft excess has a major impact on the broadband model, and does argue in favour of a more complex continuum.  

As a further test, we investigate whether a lamp-post coronal geometry can equally or better reproduce the 2016 spectra. We thus replaced the \relxillcp model with \relxilllpion, a version where the primary X-ray source is assumed to be a point source located along the rotational axis of the black hole. This model is physically and geometrically self-consistent in terms of reflection strength and disc emissivity profile, as $\mathcal{R}$ can be calculated directly from the source height in the lamp-post configuration through the parameter \texttt{FixReflFrac} (fixed to 1). Moreover, a radial ionization gradient is allowed for in the disc. We adopt an empirical power-law gradient, with $\xi$ evaluated at $R_{\rm in,disc}$ and declining as $r^{-p_{\xi}}$, where we imposed $p_{\xi}\geq0$. We find that the X-ray source height is very low, constrained within $h< 2.1\,\rg$, implying that the X-ray illumination of the disc is centrally concentrated due to severe light bending (in agreement with the steep emissivity index $q_1 > 9$ within $R_{\rm br}\sim4\,\rg$ in \relxillcp). This model prefers a maximally spinning black hole ($a>0.98$) and a smaller disc inclination ($i\sim30^{\circ}$), while iron abundance is moderately super-solar at $A_{\rm Fe}\sim2.3$. 

Interestingly, the fit converges to a flat ionization profile, with $p_{\xi}=0$ hence constant $\xi$. All the other parameters remain stable. Overall, this model returned a worse fit ($\dchidof=+45$/$+1$) compared to the \relxillcp case, with $\chis=542/410$. By relaxing the condition on the index $p_{\xi}$, the best-fit value would become $p_{\xi}\simeq-0.2$, implying that the ionization of the disc moderately increases with radius. Although the fit improves by $\Delta\chi=-15$ for the same degrees of freedom we find this solution unlikely (see also Fig. 2\ in \citealt{Kammoun19}), as it suggests (in first approximation) that the density of the disc drops faster than $r^{-2}$, which is incompatible with a standard disc. In fact, if we adopt instead in \relxilllpion the ionization profile based on the density of an $\alpha$-accretion disc \citep{Shakura73}, the fit statistics further deteriorates down to $\chis=574/411$. 

We then replaced \relxilllpion with \relxilld, where the (constant) accretion disc density is a free parameter. The reflected spectra are computed allowing for a disc density ranging from the standard case of $\log( n/\rm cm^{-3})=15$ up to $\log( n/\rm cm^{-3})=19$ \citep{Garcia16}. In a higher density regime, the thermal emission at soft X-ray energies would increase and so raise the continuum flux at $E\lesssim 2\kev$, allowing the fit of a stronger soft X-ray excess. This model is therefore particularly relevant to the case of \ton. By letting the disc density vary, however, the best-fit value remains pegged at the lower limit. We constrain $\log( n/\rm cm^{-3})<16.0$, which is consistent with the previous measurement by \citet {Jiang19}, i.e., $\log( n/\rm cm^{-3})\sim15.6$\,\cmq. The overall statistics of $\chis=491/408$, however, are slightly improved over the \relxillcp model, implying that a simple cut-off power-law shape is marginally preferred for the primary continuum.

Given that the soft ($E<3$ keV) and hard ($E>3$ keV) bands of \ton apparently require distinct reflection parameters (Table \ref{tab:broadband_relxill_no_xillver_2016}), as a final test we built a model with two \relxilld components. Specifically, we created a model where the innermost regions of the disc are allowed to have different density, emissivity index and ionisation parameter compared to the outer regions. We use the parameter $R_\rmn{br}$ to set the boundary between the inner and the outer disc. Iron abundance and reflection strength are tied between the two components, as well as the geometrical parameters (spin, inclination). In this configuration, the inner component, subject to stronger blurring effects and possibly denser, could account for the soft excess, while the hard reflection features could arise at larger distance. At this stage, the \xillver distant reflector is discarded. This dual \relxilld model turns out to be statistically similar to the \relxillcp$+$\xillver and \relxilld$+$\xillver models, with $\chis=492/405$. The inner region is suggested to be very compact, confined within the central $\sim$5\,$\rg$ (although with loose constraints, possibly due to some degeneracy between the two components), and it is characterized by nearly maximal spin, steep emissivity, and high ionization. Contrary to the expectations, however, its density remains low, $\log( n_1/\rm cm^{-3})<15.1$. The density is instead larger farther out, where $\log( n_2/\rm cm^{-3})\sim16.4$, causing a drop of the ionization to $\logxi = 0$. If anything, this result supports the presence of a reflection component that does not arise from an extreme gravity regime, which, given the shape of the Fe\,K complex, seems more likely associated with the outer disc rather than with a pc-scale reprocessor.

The best-fit parameters for all the reflection models are reported in Table\,\ref{tab:broadband_relxill_models_2016}. The corresponding data/model ratios of the \relxilllpion (top panel), \relxilld (middle) and dual \relxilld (bottom) fits are shown in \fig\ref{fig:ts180_2016_relLP_CP_D_ra}, where significant residuals are still present. It is clear that none of these versions are able to reproduce the whole 0.4--30 keV \xmmnu spectrum. In fact, we find that the lamp-post scenario, which is the most rigid due to the enforced self-consistency in terms of disc emissivity, reflection strength and possibly also ionization profile (in the $\alpha$-disc variant), provides the worse fit to the data compared to the other reflection models.

\begin{table*}

\begin{tabular}{c@{\hspace{15pt}}c@{\hspace{25pt}}c@{\hspace{25pt}}c}

\hline

Component                   &Parameter             &Best-fit value &Description\,(units) \\

\hline \\[-2ex]

%    \multirow{1}{*}{\texttt{Tbabs}}

\texttt{Tbabs}                                     &$\nhgal$     &$1.3\times10^{20}$&Galactic column (\cmsq)\\

\\
%\multirow{4}{*}{\texttt{nthComp}}

\texttt{nthComp} &$kT_{\rm bb}$             &$5.5^{+1.0}_{-1.1}$&Seed photon temperature\,(eV)\\[1ex]

                                   &$\Gamma_{\rm warm}$            &$3.02_{-0.13}^{+0.11}$&Photon index, warm corona \\[1ex]

                                   &$kT_{\rm e, warm}$             &$0.32_{-0.04}^{+0.06}$&Temperature, warm corona\,(keV)\\[1ex]

%					               &$\tau_{\rm warm}$              &$14.0_{-0.8}^{+0.8}$&Optical depth of warm corona\\

						      &norm       &$6.8_{-0.9}^{+0.9}\times10^{-4}$&Normalization (arbitrary)\\[1ex]

\\

%\multirow{9}{*}{\relxillcp}

\relxillcp                                   &$q$            &$3^{*}$&Emissivity index ($q=q_{1}=q_{2}$)\\[1ex]

%					               &$R_\rmn{br}$            &$15^{*}$&Break radius\,($\rg$)\\

					               &$a$            &$<0.70$&Black hole spin\\[1ex]

						       &$i$            &$42.2_{-5.9}^{+5.5}$&Inclination\,(degrees)\\[1ex]

					               &$\Gamma$          &$2.26_{-0.06}^{+0.06}$&Photon index, hot corona \\[1ex]

					               &$\log(\xi)$    &$<0.1$&Ionization\,(erg\,cm\,s$^{-1}$) \\[1ex]
					                     					             &$kT_{\rm e}$   &$100^*$&Temperature, hot corona (keV)\\[1ex]                
					               
					               &$A_{\rm Fe}$    &$1^{*}$&Iron abundance\,(solar)\\[1ex]

%					               &$\tau_{\rm hc}$            &$<0.72$&Optical depth of hot corona\\

					               &$\mathcal{R}$  &$0.5_{-0.2}^{+0.2}$&Reflection fraction\\[1ex]

						       &norm           &$1.6_{-0.1}^{+0.1}\times10^{-5}$&Normalization (arbitrary)\\

\\

%\multirow{1}{*}{\texttt{smallBB}}  
\texttt{smallBB} &norm           &$1.3_{-0.3}^{+0.2}\times10^{-2}$&Normalization ($\rm cm^{-2}\,s^{-1}$)\\

\\

                                  &MOS   &$1.05\pm0.01$&\multirow{3}{*}{Cross-normalization}\\

                                  &FPMA  &$1.29\pm0.05$\\

                                  &FPMB  &$1.34\pm0.05$\\

\hline

Fit statistic            &$\chi^2/\nu$   &$430.5/410 	$\\

\hline

\end{tabular}

\caption{Summary of the best-fit parameters of the phenomenological warm-corona model applied to the 2016 \xmmnu optical/UV to hard X-ray SED of \ton, which also includes a contribution from the small blue bump at $3000\,\ang$ (see text for details). $^{*}$ denotes a frozen parameter.}%was also added with normalization of $6\pm2\times10^{-3}UNITS$ (see text for details).}
\vspace{-5mm}
\label{tab:broadband_2nthc_2016}
\end{table*}

\subsection{Phenomenological warm corona: \nthcomp}
\label{subsec:Modelling with Comptonization}

Having established that relativistic reflection alone cannot self-consistently describe the 2016 \xmmnu spectra of \ton, we subsequently included the contribution of warm Comptonization. In order to model each of the Comptonized coronal emissions (warm and hot) we adopted \nthcomp. There are three main parameters that can be obtained in \nthcomp: the electron temperature of the plasma $kT_{\rm e}$, the seed disc-photon temperature $kT_{\rm bb}$ and the asymptotic power-law photon index of the Comptonized spectrum $\Gamma$. On this basis, we fitted the 2016 broadband \xmmnu spectra with a phenomenological warm corona model constructed as follows: \texttt{Tbabs}*(\texttt{smallBB}\,+\,\texttt{nthComp$_{\rm warm}$}+\,\relxillcp). The hot corona parameters are already hardwired in \relxillcp, where the incident spectrum is obtained with \nthcomp as well \citep[][see also Section\,\ref{subsec:Modelling with relativistic reflection}]{Garcia14,Dauser14}. 

For this test, we also take advantage of the U and UVM2 photometric points provided by the OM, in order to extend the analysis to the optical/UV band. By doing so, decide to leave the temperature of the `seed' photons free in this fitting procedure.\footnote{We note that this choice introduces a small internal inconsistency, as in the in \relxillcp component $kT_{\rm bb}$ is hardwired to 50 eV. However, the X-ray spectral shape is fully insensitive to exact value of $kT_{\rm bb}$, provided that it is well outside the fitted bandpass.} Moreover, we also consider the contribution from the broad line region (BLR), which is responsible for the small blue bump emission at $\sim$3000\,$\ang$ \citep{Grandi82}. We modelled this component with an additive table (\texttt{smallBB}), where the normalization is the only free parameter in units of $\rm photons\,cm^{-2}\,s^{-1}$. The physical assumption was presented in detail by \citet{Mehdipour15} for NGC\,5548. Its inclusion improved the fit by $\dchidof=-25$/$-1$, corresponding to a 5$\sigma$ confidence level. 

The warm corona component is found to have an electron temperature of $kT_{\rm e, warm}\sim320$ eV with a photon index of $\Gamma_{\rm warm}\sim3$. From these values and following the method described in \citet{Beloborodov99}\footnote{The optical depth is estimated from the following equation: $\Gamma\simeq\frac{9}{4}y^{-2/9}$, where $y=4(\Theta+4\Theta^{2})\tau(\tau + 1)$ is the Compton parameter and $\Theta=kT_{\rm e}/m_{\rm e}c^{2}$.}, we estimate the corresponding optical depth to be $\tau_{\rm warm}=9.8^{+2.5}_{-1.8}$. This can be identified with the extended and optically thick plasma that covers a large fraction of a `passive' accretion disc in the models of \citet{Rosanska15} and \citet{Petrucci18}. Indeed, no direct disc emission is required. With this respect, it is worth noting that $kT_{\rm bb}\simeq5.5$ eV, which is significantly lower than predicted for a standard disc and a black hole with mass $\mbh\sim10^{7}\Msun$ accreting at the Eddington limit in \ton \citep[e.g.,][]{Turner02}. Values of $kT_{\rm bb}$ of a few eV, irrespective of $\mbh$, are systematically found in the `passive disc' scenario \citep[e.g.,][]{Petrucci18}, and seem to be required to fit the OM data. We stress, however, that all the other model parameters, which are entirely driven by the X-ray spectra, are not affected by $kT_{\rm bb}$.

Here, we chose \relxillcp to model simultaneously the hot corona responsible for the primary continuum and the reflection features as the broad ($\sigma\sim300$ eV) \iron emission line at $E_{\rm rest}\sim6.6$ keV and the X-ray emission above 10 keV. For the hot corona, we find a lower limit on the electron temperature of $kT_{\rm e, hot}>60$ keV, which was then fixed to 100 keV. With $\Gamma_{\rm hot}\simeq2.26$, this corresponds to $\tau_{\rm hot}\simeq0.5$. These values are consistent with the standard optically thin and hot plasma that shapes the primary X-ray continuum in AGN \citep[e.g.,][]{Fabian15}. During the fitting procedure the disc emissivity is assumed to be at its default classical solution, i.e., $q=3$. The reflection fraction is rather low, measured at $\mathcal{R}=0.5\pm0.2$, and an upper limit on the spin value of $a<0.70$ is preferred in the fit, where a maximal spin is ruled out at the $\gtrsim3\sigma$ confidence level. The inclination is about $\sim40^{\circ}$. Interestingly, these results are nearly identical to the findings of the \relxill fit over the 3--30 keV range, and are also consistent with the 2015 \xmm sequence reported in \citet{Parker18ton}.

\fig\ref{fig:ts180_2016_2nth_eeuf_ra.pdf}\,(top) shows the best fitting warm-corona model (red) for the 2016 \xmmnu observation, where the individual contributions of the warm corona (orange), hot corona (magenta) and mild relativistic reflector (cyan) are also displayed. The warm corona scenario explains very well the \xmmnu spectra, and it statistically yields a very good fit with $\chis=430/410$, which is a considerable improvement over any of the above relativistic reflection models. The best-fit values are reported in Table\,\ref{tab:broadband_2nthc_2016}. It should be kept in mind, however, that this model is purely phenomenological, and that its construction does not ensure a physical self-consistency.

\subsection{Self-consistent warm corona: \optxagn}
\label{subsec:Broadband modelling with optxagn}

\begin{table*}

\begin{tabular}{c@{\hspace{15pt}}c@{\hspace{25pt}}c@{\hspace{25pt}}c}

\hline

Component                   &Parameter             &Best-fit value &Description\,(units) \\

\hline\\[-2ex]

%    \multirow{1}{*}{\texttt{Tbabs}}

\texttt{Tbabs}                                     &$\nhgal$     &$1.3\times10^{20}$&Galactic column (\cmsq)\\

\\

%\multirow{10}{*}{\texttt{optxagnf}}

\texttt{optxagnf}				   &$\mbh$                &$(4.0\times10^{7})^*$ &Black hole mass ($\Msun$) \\[1ex]

				   &$D$          &$262.1^{*}$&Co-moving proper distance (Mpc) \\[1ex]

                                   &$a$                &$0^*$ & Black hole spin \\[1ex]

                                   &$\Gamma_{\rm hot}$                   &$2.26_{-0.06}^{+0.05}$&Photon index, hot corona \\[1ex]

				   &$kT_{\rm e,hot}$              &$100^{\rm df}$&Temperature, hot corona\,(keV)\\[1ex]

                                   &$kT_{\rm e,warm}$              &$0.31_{-0.04}^{+0.06}$&Temperature, warm corona\,(keV)\\[1ex]

					               &$\tau_{\rm warm}$                       &$11.2_{-1.4}^{+1.5}$&Optical depth, warm corona\\[1ex]

				   &log$(\leddratio)$                &$-0.34_{-0.08}^{+0.08}$&Eddington ratio\\[1ex]

                                   &$R_{\rm cor}$           &$10.1_{-0.3}^{+0.4}$&Coronal radius\,($\rg$)\\[1ex]

&$f_{\rm pl}$           &$0.40_{-0.05}^{+0.05}$&Energy dissipated in the hot corona\,(\%)\\[1ex]

%					              &norm                          &$1.5^{\rm t}$&Normalization corrected for inclination effect \\

\\

%\multirow{7}{*}{\relxillcp}

\relxillcp                                   &$q$            &$3^{*}$&Emissivity index ($q=q_{1}=q_{2}$)\\[1ex]

%					               &$R_\rmn{br}$            &$15^{*}$&Break radius\,($\rg$)\\

%					               &$a$            &$0^{\rm t}$&Black hole spin tied to $a$ in \optxagn \\

					               &$i$            &$41.8_{-9.0}^{+7.0}$&Inclination\,(degrees)\\[1ex]

								   &$R_{\rm in,refl}$            &$10.1^{\rm t}$&Reflection inner radius ($\rg$) -- tied to $R_{\rm cor}$\\[1ex]

					               &$\Gamma$       &$2.26^{\rm t}$&Photon index -- tied to $\Gamma_{\rm hot}$ \\[1ex]

					               &$\log(\xi)$    &$<0.1$&Ionization\,(erg\,cm\,s$^{-1}$) \\[1ex]
					                     					                     
					               &$A_{\rm Fe}$    &$1^{*}$&Iron abundance\,(solar)\\[1ex]

%					               &$kT_{\rm e}$   &$100^{\rm t}$&Temperature of hot electrons fixed at the \optxagn value (keV)\\
					                     
%					               &$\mathcal{R}$  &$-1^{*}$&Reflection fraction\\

						       &norm           &$8.3_{-2.6}^{+3.0}\times10^{-6}$&Normalization (arbitrary)\\

\\

%\multirow{1}{*}{\texttt{smallBB}}  
\texttt{smallBB} &norm           &$6.7_{-2.0}^{+1.9}\times10^{-3}$&Normalization ($\rm cm^{-2}\,s^{-1}$)\\

\\

                                  &MOS   &$1.05\pm0.01$&\multirow{3}{*}{Cross-normalization}\\

                                  &FPMA  &$1.29\pm0.05$\\

                                  &FPMB  &$1.34\pm0.05$\\

\hline

Fit statistic            &$\chi^2/\nu$   &$428.0/411$\\

\hline

\end{tabular}

\caption{Summary of the best-fit parameters of the \optxagn model applied to the 2016 \xmmnu optical/UV to hard X-ray SED of \ton. $^{\rm t}$ and $^{*}$ respectively denote tied and frozen parameters during fitting, while $^{\rm df}$ indicates a fixed setting hardwired into a given model component. For the reflection component, spin and temperature of the hot corona are tied to the corresponding values in \optxagn. We further assume that the reflection inner radius, $R_{\rm in,refl}$, extends to the radius of the corona, $R_{\rm cor}$ (see text). A contribution from the small blue bump at $3000\,\ang$ was also added.}
\vspace{-5mm}
\label{tab:broadband_optx_2016}
\end{table*}

Moving forward with the UV/X-ray SED fitting procedure, we replaced the phenomenological warm corona model (i.e., \nthcomp) with the physically motivated \optxagn \citep{Done12}. This assumes a similar physical scenario but with the warm corona embedded in the disc and energetically self-consistent. The overall model is constructed as follows: \texttt{Tbabs}*(\texttt{smallBB}+\optxagn+\relxillcp). The \optxagn model consists of three main spectral components which are powered by dissipation in the accretion flow. The first contribution is from the thermal optical/UV emission arising from the outer regions of the accretion disc; the second component is produced via Compton up-scattering of `seed' UV photons into soft X-ray photons in a warm corona; the third component is the primary X-ray power-law emission from Comptonization in the hot corona. Compared to the previous model, in \optxagn it is possible to measure the coronal size ($R_{\rm cor}$), defined as a transitional radius from the outer disc to the inner Comptonizing region. The parameter $f_{\rm pl}$ is the fraction of the total energy dissipated in the hot corona and emerging as the hard power-law component, while the remaining $1-f_{\rm pl}$ fraction is dissipated in the warm corona as the soft excess. The parameter $\log(L_{\rm bol}/\ledd)\equiv \log(\dot{M}/\dot{M}_{\rm Edd} )$ is the Eddington ratio. The parameters $kT_{\rm e}$ and $\tau$ are the electron temperature and optical depth of the warm corona, respectively. The calculation of the \optxagn model assumes a disc inclination angle of $60^{\circ}$, when the normalization is fixed to unity. In order to take into account the inclination effects on the \optxagn emission, we tied the normalization to the inclination angle of \relxillcp using the relation $\cos(i)/\cos(60^{\circ})$.

We find that the temperature and optical depth of the warm corona component are broadly consistent with the phenomenological model at $kT_{\rm e,warm}\simeq310$ eV and $\tau_{\rm warm}\sim11$. The fraction of energy released in the hot corona is $f_{\rm pl}\sim40$ per cent. The coronal size is reasonably compact at $R_{\rm cor}=10.1_{-0.3}^{+0.4}\,\rg$, which would explain also the presence of a moderately broad \iron component. In fact, we have assumed here that disc reflection arises at distances larger than $R_{\rm cor}$, by tying $R_{\rm in,refl} = R_{\rm cor}$. Note that the physical inner radius of the disc ($R_{\rm in,disc} = R_{\rm isco}$) cannot be assessed through reflection in this configuration, hence we fixed the spin to $a=0$. Interestingly, the Eddington ratio is measured at $\log(\leddratio)=-0.34\pm0.08$, which implies that \ton radiates at a considerable fraction (at least $\sim50$ per cent) of its Eddington luminosity.

In \ton, the black hole mass is still rather uncertain. From the optical spectrum obtained at the ESO 1.52-m telescope (La Silla) in 1996 October, presented in \citet{Comastri98}, we obtain a single-epoch virial $\mbh=9.5_{-2.9}^{+3.6}\times10^{6}\,\Msun$ using the \citet{Bentz13} relation with virial factor $f_{\rm vir} = 4.3\pm1.1$. When fixing $\mbh$ in \optxagn to the face-value of $\mbh=10^{7}\,\Msun$, the model either struggles to account for the UV to soft X-ray band or returns values of $R_{\rm cor}$ of several hundreds of $\rg$, which are unphysically large under the assumption that the disc is not completely passive (see Section\,\ref{subsec:Modelling with Comptonization}). We therefore allowed $\mbh$ to vary and then froze it to the preferred value of $\mbh \sim 4\times10^{7}\,\Msun$. The Eddington ratio inferred above should therefore be considered as a conservative lower limit. In all the SED fits presented from here on, we still included the contribution of the small blue bump from the BLR and the usual reflection component, as shown in \fig\ref{fig:ts180_2016_2nth_eeuf_ra.pdf} (middle panel). This physically motivated warm corona scenario returned an excellent fit to the data, i.e., $\chidof=428/411$. The best-fit values are tabulated in Table\,\ref{tab:broadband_optx_2016}.

\begin{table*}

\begin{tabular}{c@{\hspace{15pt}}c@{\hspace{25pt}}c@{\hspace{25pt}}c}

\hline

	Component                   &Parameter             &Best-fit value &Description\,(units) \\

\hline\\[-2ex]

%    \multirow{1}{*}{\texttt{Tbabs}}

\texttt{Tbabs}                                     &$\nhgal$     &$1.3\times10^{20}$&Galactic column (\cmsq)\\

\\

%\multirow{10}{*}{\agnslim}

\agnslim				   &$\mbh$                &$(1.0\times10^{7})^{*}$ &Black hole mass ($\Msun$) \\[1ex]

				   &$D$          &$262.1^{*}$&Co-moving proper distance (Mpc) \\[1ex]

				   &$a$                &$<0.37$&Black hole spin \\[1ex]

%                               &$\cos(i)$   &$(0.74_{-0.07}^{+0.05})^{\rm t}$&cosine of the inclination angle tied to $i$ of \relxillcp as $\cos(42.1_{-5.7}^{+4.4})$\\

                                   &$kT_{\rm e, hot}$              &$100^*$&Temperature, hot corona\,(keV)\\[1ex]

                                   &$\Gamma_{\rm hot}$                   &$2.28^{+0.06}_{-0.06}$&Photon index, hot corona\\[1ex]

&$R_{\rm hot}$           &$5.5^{+1.1}_{-0.1}$&Outer radius, hot corona\,($\rg$)\\[1ex]

 &$kT_{\rm e, warm}$              &$0.29^{+0.05}_{-0.03}$&Temperature, warm corona\,(keV)\\[1ex]

	&$\Gamma_{\rm warm}$                  &$2.89^{+0.24}_{-0.14}$&Photon index, warm corona \\[1ex]

                                   &$R_{\rm warm}$           &$<8.8$&Outer radius, warm corona\,($\rg$)\\[1ex]

				                                      &$\logmdot$                &$0.67_{-0.14}^{+0.06}$&Eddington ratio\\[1ex] %, where $\logmdot\equiv\log(\leddratio)$ 

\\

%\multirow{7}{*}{\relxillcp}

\relxillcp                                   &$q$            &$3^{*}$&Emissivity index ($q=q_{1}=q_{2}$)\\[1ex]

%					               &$R_\rmn{br}$            &$15^{*}$&Break radius\,($\rg$)\\

%						       &$a$            &$<0.35^{\rm t}$&Black hole spin\\

					               &$i$            &$42.4_{-5.0}^{+5.3}$&Inclination\,(degrees)\\[1ex]

								   &$R_{\rm in,refl}$            &$8.8^{\rm t}$&Reflection inner radius ($\rg$) -- tied to $R_{\rm warm}$\\[1ex]

					               &$\Gamma$       &$2.28^{\rm t}$&Photon index -- tied to $\Gamma_{\rm hot}$\\[1ex]

						       &$\log(\xi)$    &$<0.1$&Ionization\,(erg\,cm\,s$^{-1}$) \\[1ex]
					                     					                     
					               &$A_{\rm Fe}$    &$1^{*}$&Iron abundance\,(solar)\\[1ex]

%					               &$kT_{\rm e}$\,(keV)   &$>47^{\rm t}$&Temperature of hot electrons\\
					                     
%					               &$\mathcal{R}$  &$-1^{*}$&Reflection fraction\\

						       &norm           &$8.9_{-3.0}^{+3.0}\times10^{-6}$&Normalization (arbitrary)\\

\\

%\multirow{1}{*}{\texttt{smallBB}}  
\texttt{smallBB} &norm           &$1.6_{-0.2}^{+0.2}\times10^{-2}$&Normalization ($\rm cm^{-2}\,s^{-1}$)\\

\\

                                  &MOS   &$1.05\pm0.01$&\multirow{3}{*}{Cross-normalization}\\

                                  &FPMA  &$1.29\pm0.04$\\

                                  &FPMB  &$1.33\pm0.05$\\

\\

								&0.5--2 keV &$4.80_{-0.04}^{+0.03}\times10^{-12}$& \\  [1ex]

Flux								&2--10 keV &$3.05_{-0.07}^{+0.07}\times10^{-12}$ & Observed fluxes\,(erg\,cm$^{-2}$\,s$^{-1}$)\\[1ex]

								&10--30 keV  &$2.26_{-0.08}^{+0.09}\times10^{-12}$ & \\

\hline

Fit statistic            &$\chi^2/\nu$   &$428.2/410$\\

\hline

\end{tabular}

\caption{Summary of the best-fit parameters of the \agnslim model applied to the 2016 \xmmnu optical/UV to hard X-ray SED of \ton. The disc inner radius $(R_{\rm in,disc})$ is self-consistently determined by the model and, for this set of parameters, coincides with $(R_{\rm isco})$. $^{\rm t}$ and $^{*}$ respectively denote tied and frozen parameters during fitting. For the reflection component, the spin is tied to the corresponding value in \agnslim, and we assume that $R_{\rm in,refl}$ extends down to the radius of the warm corona, $R_{\rm warm}$. A contribution from the small blue bump at $3000\,\ang$ was also added.}
\vspace{-5mm}
\label{tab:broadband_agnslim_2016}
\end{table*}

\subsection{Warm Comptonization at high accretion rates: \agnslim}
\label{subsec: two-corona broadband modelling with AGNSLIM}

As \ton is accreting close to its Eddington rate, we also model the 2016 UV/X-ray SED with \agnslim. This spectral library was developed by \citet{Kubota18,Kubota19} for super-Eddington black hole accretion, based on the emissivity model from a radially stratified disc \citep{Abramowicz88}. The radial advection keeps locally the surface luminosity at the Eddington limit, resulting in a radial emissivity profile $L(r) \propto r^{-2}$ throughout the disc, where $L(r)$ is the surface luminosity as a function of radius. As in \optxagn, there are three distinct emitting regions: the outer disc from $R_{\rm out}$ to $R_{\rm warm}$; a warm Comptonizing region, between $R_{\rm warm}$ and $R_{\rm hot}$, producing the soft X-ray excess; an inner hot Comptonizing region from $R_{\rm hot}$ extending, untruncated, down to the inner radius of the flow.

The model is constructed by simply replacing \optxagn with \agnslim model, thus obtaining: \texttt{Tbabs}*(\texttt{smallBB}\,+\,\agnslim +\,\relxillcp). The inclination angle of the disc is variable through the parameter $\cos(i)$, as opposed to the fixed value of 60$^{\circ}$ in \optxagn. In the fitting procedure we tied it to the inclination parameter of \relxillcp, for a viewing angle of $i=42.4_{-5.0}^{+5.2}$ deg.

The \agnslim parameters measured for the warm Comptonizing component are largely consistent with the previous results: the electron temperature is $kT_{\rm e,warm}\simeq290$ eV, while $\Gamma_{\rm warm}\sim2.9$ is fully consistent with the RGS spectral slope. Differently from the previous model, we allowed the black hole spin to vary as the exact position of the ISCO is important in \agnslim. The best-fit value is now $a\simeq0.34$, but this is again consistent with zero at the 90 per cent confidence level ($a<0.37$).

\begin{figure*}
  \includegraphics[width=0.68\textwidth]{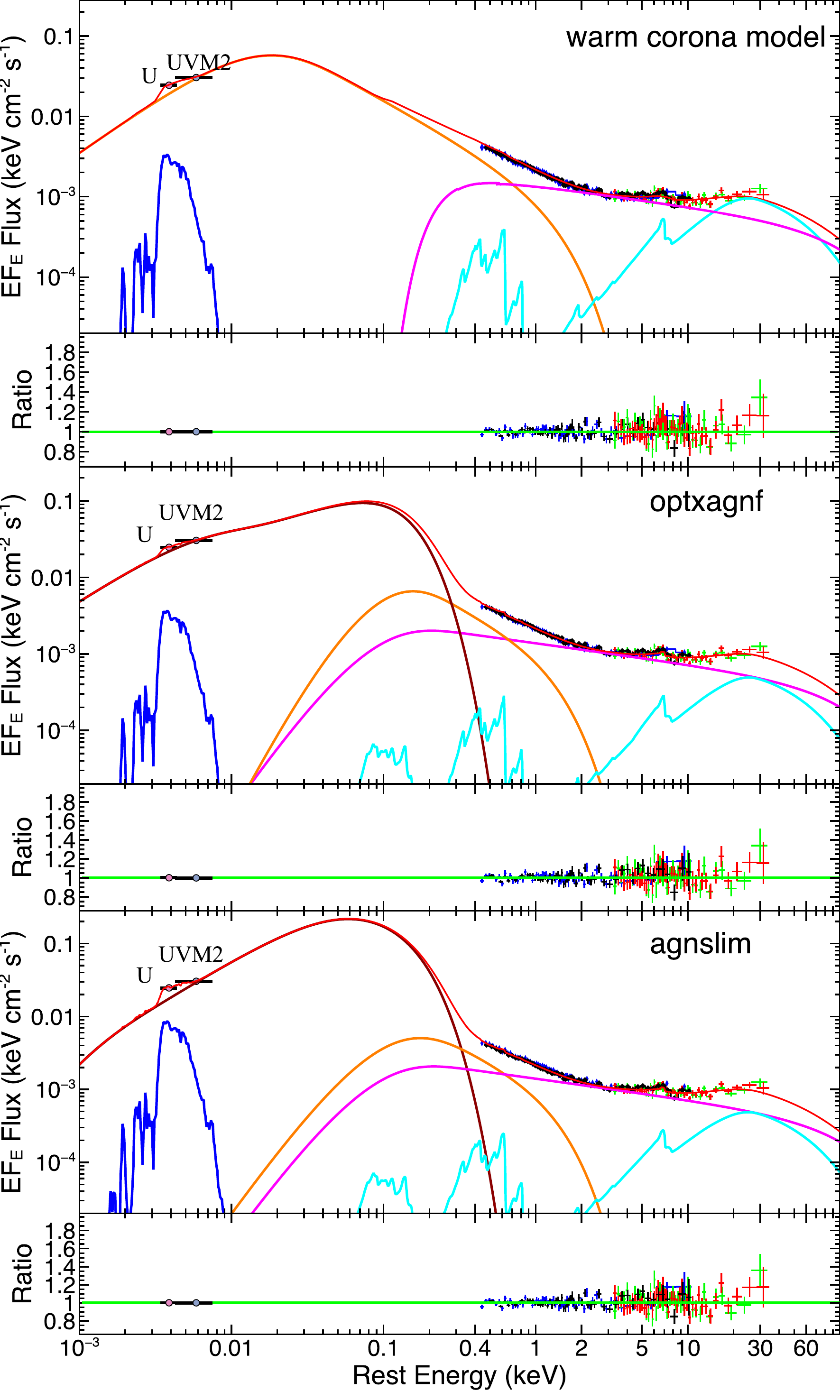}
  %\vspace{-50mm}
	\caption{Top panel: optical to hard X-ray SED of \ton, where the best-fitting phenomenological warm-corona model (red solid line) is superimposed to the 2016 \xmmnu `fluxed' spectra (unfolded against a power law with $\Gamma=2$). The EPIC pn, MOS\,1$+$2, and FPMA/B spectra are shown in black, blue, red and green, respectively. The OM photometric U and UVM2 points are also included in order to better constrain the properties the warm corona and the spectral shape of the Comptonized component (orange). The other contributions to the observed spectrum come from the primary continuum (due to Comptonization in the hot corona, plotted in magenta), the relativistic reflection accounting for the broad \fe emission and Compton hump above 10 keV (cyan), the thermal disc emission (brown, absent in the phenomenological model), and the small blue bump emission from the BLR at $\sim3000\,\ang$ (blue). Middle panel: as above for the \optxagn model. Bottom panel: same, where \optxagn is replaced with the super-Eddington \agnslim model. The data/model ratios corresponding to each fit are shown in the panels below the spectra. All the X-ray data and models are corrected for a Galactic absorption column of $\nhgal=1.3\times10^{20}$\,\cmsq, and that the OM data have been de-reddened as described in Section\,\ref{sec:Observations and Data Reduction}. The luminosity between 1 eV and 100 keV is estimated to be $L_{\rm bol}\approx 2$--$6 \times 10^{45}\,\ergs$ in the three models, the lower luminosity corresponding to the phenomenological model. This is due to the fact that in this scenario the disc is completely passive, and consequently there is no big blue bump in the extreme UV (i.e., the SED peak around 60--80 eV).}
\label{fig:ts180_2016_2nth_eeuf_ra.pdf}
\end{figure*}

The outer radii of the hot and warm emitting regions are estimated to be $R_{\rm hot}=5.5_{-0.1}^{+1.1}\,\rg$ and $R_{\rm warm}<8.8\,\rg$ (but always larger than $R_{\rm hot}$, see below), respectively. These measurements confirm that the warm plus hot Comptonizing regions are rather compact. This is largely consistent with \optxagn, despite the different geometry and inferred BH mass (see Section\,\ref{subsub:Combining relativistic reflection and Comptonization to estimate the size-scale of the warm corona}). For the hot Comptonizing region, the electron temperature is $kT_{\rm e,hot}>47\,\kev$ (then again fixed to 100 keV), whereas the photon index is $\Gamma_{\rm hot}=2.28\pm0.06$. This continuum slope is also assumed for the reflection component, as leaving $\Gamma_{\rm refl}$ free gives a consistent value for a marginal statistical improvement ($\Delta\chi^2>-1$).
Here the Eddington ratio $\dot m \equiv \logMdot \equiv\leddratio$ is constrained to lie in the super-Eddington regime, $\logmdot=0.67_{-0.14}^{+0.06}$. Overall, the \agnslim model provided an excellent fit to the 2016 data, with $\chis=428/410$ (see Table\,\ref{tab:broadband_agnslim_2016}). The best-fit \agnslim model, with the corresponding residuals, is shown in the bottom panel of \fig\ref{fig:ts180_2016_2nth_eeuf_ra.pdf}.

\begin{figure*}
  \includegraphics[width=0.95\textwidth]{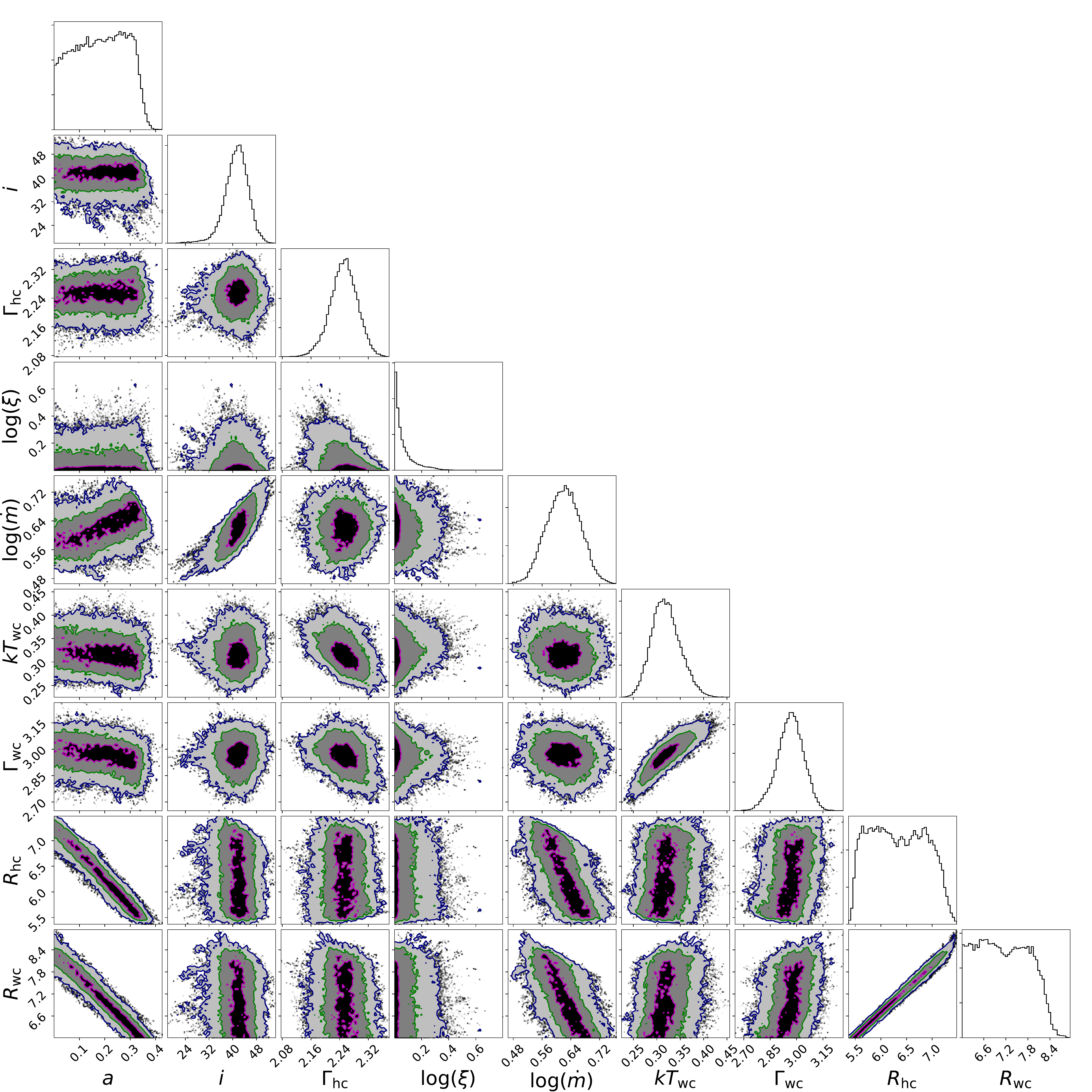}
  %\vspace{-50mm}
	\caption{MCMC contours computed for the \agnslim model applied to the 2016 \xmmnu spectra. The confidence contours of the 1 (magenta), 2 (green) and 3$\sigma$ (blue) levels are shown. A clear degeneracy is visible between some of the parameters, most notably $R_{\rm hot}$, $R_{\rm warm}$ and $a$, and, to a milder degree, the disc inclination $i$ and the accretion rate $\dot m$. These degeneracies have all a natural physical explanation (see text for details.)}
\label{fig:ts180_2016_agnslim_NS5OS3_MCMC_500kst_100kbu}
\end{figure*}

As a final test, given the model complexity, we searched for possible degeneracies by adopting the \xspecmcmc\footnote{\url{https://github.com/jeremysanders/xspec_emcee}}$^,$\footnote{\url{https://github.com/zoghbi-a/xspec_emcee}} implementation of the \texttt{emcee} code \citep{Foreman-Mackey13}. The MCMC contours for the best-fit \agnslim model (\fig\ref{fig:ts180_2016_agnslim_NS5OS3_MCMC_500kst_100kbu}) are calculated from 500,000 points, using 100 walkers and burning the first 100,000 steps. 
\fig\ref{fig:ts180_2016_agnslim_NS5OS3_MCMC_500kst_100kbu} shows some mild degeneracies between the key coronal parameters. For instance, $R_{\rm hot}$ and $R_{\rm warm}$ are strongly correlated with each other, which ensures that $R_{\rm warm}>R_{\rm hot}$, while both are unsurprisingly anti-correlated with the spin (the higher the spin, the smaller $R_{\rm isco}$). Also the inclination and the accretion rate are naturally linked to each other: the disc is inherently an anisotropic emitter of optical/UV radiation, hence for a given observed luminosity the higher the inclination, the higher the accretion rate \citep[e.g.,][]{DavisLaor11}. 
As mentioned, the temperature of the hot corona ($kT_{\rm hot}$) is poorly constrained, as is iron abundance (which was eventually frozen to solar). A stringent upper limit is measured for the ionization parameter of the disc, $\logxi<0.2$, suggesting that the reflection component arises at rather large distance. Such a case was also found by \citet{Porquet18,Porquet19} whilst fitting the broadband spectra of \arkone with a warm/hot-coronal model. 

This notwithstanding, we caution against taking such a low ionization at face value. Indeed, all the other reflection properties are consistent with those obtained over the 3--30 keV band with a plain \relxill model (Table\,\ref{tab:broadband_relxill_no_xillver_2016}). Since the observed prominence of a feature like the \feka depends on the ionization, the $\logxi<0.2$ value apparently compensates for the fact that the primary continuum in this model is described by a different component (i.e., \agnslim). This is therefore a possible shortcoming associated with the imperfect self-consistency of `hybrid' reflection plus warm Comptonization models.

\section{Discussion}
\label{sec:Discussion}

We presented a detailed broadband spectral analysis of the 0.4--30 keV X-ray emission of the NLSy\,1 galaxy \ton, simultaneously observed by \xmmnu for the first time. The source was caught in a low flux state compared to the three previous \xmm observations. However, as the joint coverage is only carried out towards the end of the observation, the average \nustar flux level is about $\sim30$ per cent higher than for \xmm. Thanks to the simultaneous broadband coverage, we aim to test whether the soft excess in \ton can be explained with either (i) relativistic reflection, or (ii) a warm corona scenario. To solve the problem of the soft excess, in fact, it is fundamental to have spectral coverage also above $10\kev$ \citep[e.g.,][]{Boissay16}.

\subsection{The role of relativistic reflection}

Regarding case (i), we adopt four different flavours of the \relxill package to test different coronal geometries, accretion disc densities or shape of the primary spectrum (see Section\,\ref{subsec:Modelling with relativistic reflection}). We find that each of these relativistic reflection models cannot self-consistently reproduce simultaneously the soft X-ray excess, the mildly relativistic \fe emission line, and the hard X-ray spectrum above 10 keV in \ton. A comprehensive spectral analysis using \relxilld, to allow for reflection from high-density discs, was carried out by \citet{Jiang19} on 17 Seyfert\,1 galaxies from the \xmm archive. \ton was also included in their sample and the best-fitting model favoured a disc density of $\log( n/\rm cm^{-3})\sim15.6$, successfully accounting for both the steep soft excess and the broad \fe emission line in the (stacked 2000--2016) EPIC-pn spectrum of \ton (see their Fig. 6). Interestingly, the required enhancement in the disc density is only moderate, about $4$ times larger than in the standard reflection codes (where the disc density is fixed), which had already been shown to work relatively well for \ton below 10 keV \citep{Nardini12,Walton13}.

The difference between \citet{Jiang19} and our results can be clearly attributed to the inclusion of the \nustar data (see \fig\ref{fig:ts180_2016_relxill_only}, right). The major problem in fitting the broadband X-ray emission, between 0.4--30 keV, arises in the attempt for the pure reflection models to fit such a steep and smooth soft excess. In fact, in all the permutations that we have tested in this paper, the key reflection parameters are somewhat fine-tuned to ensure extreme `blurring', combining a maximally rotating black hole ($a\gtrsim0.9$), a very steep disc emissivity profile ($q\gtrsim9$), and/or a low source height ($h<2.1\,\rg$), whilst largely missing the hard X-ray band. Indeed, when fitting the 3--30 keV \xmmnu spectra the reflection model accounted very well for both the hard X-rays and the broad \fe emission. The above issue arises as soon as we include the data below 3 keV. Notably, the key reflection parameters inferred from the 3--30 keV spectra remain remarkably stable when the soft band is reproduced through a different (i.e., Comptonized) component, which clearly supports the presumption of an alternative explanation for the soft X-ray excess.

Similar shortcomings were also encountered in a recent work focusing on another prototypical `bare' AGN, \arkone, based on a large simultaneous \xmmnu campaign carried out in 2014 \citep{Porquet18,Porquet19}. It was found that in order to reproduce the soft excess, the reflection scenario required an high degree of blurring, a maximally rotating black hole and high reflection fraction when fitting the whole 0.3--79 keV spectrum. As per \ton, the relativistic reflection models failed to account for both the soft and hard X-ray bands simultaneously in \arkone \citep[][see their Fig.\,8]{Porquet18}. Another evidence that relativistic reflection cannot easily explain the soft X-ray excess in \ton is provided by the high S/N stacked RGS spectrum. We find that the soft X-ray spectrum only shows marginal signatures of emission lines. In fact, the RGS spectrum of \ton is largely consistent with a simple power law (see \fig\ref{fig:rgs_spectra}). Given its nearly featureless nature, it is difficult to reconcile the RGS with a blurred reflection model, as in the latter case some relativistically broadened emission, such as from Fe\,L, could still be present in the spectrum \citep[e.g.,][]{Fabian09}.

For completeness, we also fitted the 2016 \xmmnu spectra with alternative relativistic reflection models from the \textsc{reflkerr}\footnote{https://users.camk.edu.pl/mitsza/reflkerr/} package \citep{Niedzwiecky19refkerr}. A detailed comparison between \relxill and \textsc{reflkerr} is beyond the scope of this work. Here we only compare the lamp-post and the high-density models, i.e., \texttt{reflkerrExp\_lp} and \texttt{reflkerrExpD}, respectively (the suffix \texttt{Exp} indicates that in both models the primary continuum takes the shape of a power law with exponential cutoff). We find that the results returned from the above reflection models are largely consistent, in terms of the reduced $\chisq$, with \relxilllpion and \relxilld. These again require extreme blurring values, moderate accretion disc density and low disc inclination, but leave significant residuals in the \fe and hard X-ray bands.

\subsection{The warm corona scenario}

In contrast, in case (ii), we find that the 0.4--30 keV band (and consistently down to the available optical/UV photometric points) can be explained by a combination of: (a) direct thermal emission from the accretion disc and/or (b) optical/UV seed disc-photons that are Compton up-scattered by a warm ($kT_{\rm e}\sim0.3$ keV) and optically thick ($\tau\sim10$) plasma, dominating below $\sim1\kev$; (c) a hot ($kT_{\rm e}>60$ keV) and optically thin ($\tau<0.5$) Comptonizing corona responsible for the primary X-ray emission above $\sim$2 keV; and (d) a mildly relativistic reflection component responsible for the \fe feature at $\sim$6.6 keV and the hard X-ray emission. 

This `hybrid' warm corona plus reflection model can simultaneously account for the soft and hard X-ray spectrum of \ton. In this scenario, hard X-ray photons are produced in the hot coronal region, where a fraction are directly observed as the power-law continuum and some would heat the upper layers of the accretion disc creating an optically thick warm coronal region. Subsequently, the UV disc photons are Comptonized into the observed soft X-ray photons.

\subsubsection{Are optically thick/warm coronae physically plausible?}
\label{subsub:Are optically thick/warm coronae physically plausible?} 

Some of the physical properties of the accretion disc/coronal systems, i.e., their exact geometry and relative location, are still largely unknown. In particular, the plausibility of a warm corona is still the subject of open debate. Recently, in a detailed analysis performed on the bright Seyfert\,1 galaxy Mrk\,509, \citet{Garcia19} found that the soft excess could be well explained with either a warm corona component or a relativistically blurred, high-density reflection model (i.e., \relxilld with $n=10^{19}$\,\cmq). Since on statistical ground neither of these scenarios prevailed, \citet{Garcia19} discussed the physical implications of the two models. It was argued that a low-temperature warm corona ($kT_{\rm e}\sim0.5$--$1$ keV) with large optical depth ($\tau\sim10$--$20$) is somewhat conflicting with the standard view of a corona where electron scattering is the main cause of opacity. In their simulations, the authors found that in such a low-temperature/optically-thick regime, atomic opacity dominates over Thomson opacity. This scenario would lead to the inevitable presence of a wealth of absorption features imprinted in the soft X-ray spectrum (see their Figs. 5 and 6), in disagreement with the smooth, featureless soft excess frequently observed in AGN (like in \ton).

On the other hand, \citet{Petrucci20} performed new simulations in order to investigate the physical and radiative properties of optically thick warm coronae. The novel ingredient in these simulations was to include the contribution of internal heating power for different sets of $\tau$ and $n$. Although the exact physical mechanisms responsible for such internal heating are unknown (but the same holds for the hot corona; e.g., \citealt[and references therein]{LussoRisaliti17}), by assuming an intrinsic heating mechanism then Comptonization is recovered as the dominant cooling process in a large part of the parameter space, where the spectral properties of the warm corona, such as $\Gamma$ and $kT_{\rm e}$, are reconciled with those observed for the soft X-ray excess. In this regime, the presence of a featureless soft excess, like in \ton, can be expected from reprocessing in a warm corona.

By comparing the simulations by \citet[][see their Figs. 2 and 3]{Petrucci20} with the measured values from our warm-corona models for temperature, optical depth and spectral shape (see Tables\,\ref{tab:broadband_2nthc_2016}, \ref{tab:broadband_optx_2016} and \ref{tab:broadband_agnslim_2016}), our results are compatible with an accretion disc which is either weakly dissipative and partially blanketed by warm plasma (\citealt{Petrucci13}, see their Fig. 10) or vertically/radially stratified with distinct Comptonizing regions (\citealt{Done12}, see their Fig. 5; \citealt{Kubota19}, see their Fig. 7). At this stage the exact geometry is basically unknown, since the subtle differences between the various models (e.g., \optxagn versus \agnslim) are hardly distinguishable in the spectral analysis. A completely passive disc seems unlikely, as this scenario involves an apparent fine-tuning of the `seed' photon temperature (see Table C.1 of \citealt{Petrucci18}), possibly driven by the specific SED coverage provided by the \xmm/OM filters. Moreover, the lack of a big blue bump would imply a much smaller AGN bolometric luminosity (Fig.\,\ref{fig:ts180_2016_2nth_eeuf_ra.pdf}). Hence,
 any configuration must allow part of the disc emission to escape without crossing the warm corona, as suggested by the presence of reflected emission features (such as the broad \fe line). These might arise from the `uncovered' mid/outer regions rather than from the innermost disc, as discussed in the next Section. 

Remarkably, the reflection component in the \agnslim model suggests that the disc is irradiated by the same power-law continuum as the one arising from the hot corona. The relative photon indices, when both are allowed to vary, are $\Gamma_{\rm refl}\simeq2.21\pm0.13$ and $\Gamma_{\rm hot}\simeq2.30\pm0.06$. This is in agreement with the generally accepted idea of a hot corona located above the disc, although in \agnslim only the radial extent of this component can be formally controlled, but not its height.

\subsubsection{The radial extent of the warm corona}
\label{subsub:Combining relativistic reflection and Comptonization to estimate the size-scale of the warm corona}

As mentioned in Section\,\ref{subsec:Modelling with Comptonization}, even in the warm-corona models a reflection component is always required to fit the broad \fe emission feature and the mild hard X-ray excess. The presence of such features suggests that any optically thick corona must have a low covering factor, so that some reflection from the accretion disc can be directly exposed to our line of sight. Following the parallel with the 2014 \xmmnu campaign on \arkone, it seems plausible that the warm corona is radially confined and/or patchy. In \arkone, the detected \fe emission line complex and the broadband reflection component \citep{Nardini16,Porquet18} apparently arise from a few tens of gravitational radii from the central black hole, i.e. outside the extent of the putative warm corona. By contrast, \citet{Matt14} detected neither a broad \fe profile nor a reflected continuum component in the 2013 \xmmnu spectra, which could be explained by a more extended corona covering the surface of the accretion disc out to a larger radius. Indeed, a possible change in coronal size was suggested by \citep{Porquet19}, where both the 2013 and 2014 broadband \xmm (OM and pn) and \nustar spectra were successfully fitted with the \optxconv model \citep{Done13}.\footnote{ This model is constructed by convolving \optxagn with the relativistic blurring calculated by the \texttt{relconv} model \citep{Dauser10}.} It was found at the 5.5$\sigma$ confidence level that the size of the warm and hot corona (which are radially co-spatial in \optxagn) decreased between the 2013 and 2014 observations, from $R_{\rm cor}=85_{-10}^{+13}\,\rg$ to $R_{\rm cor}=14\pm3\,\rg$. 

A similar analysis in \ton is precluded by the different quality (due to different duration and/or flux state) of the four \xmm observations, and by the availablity of a single-epoch \nustar spectrum. However, we performed a further test by allowing the reflection inner radius ($R_{\rm in,refl}$) in \relxillcp to vary in both the phenomenological warm corona and \optxagn models. Having fixed $R_{\rm in,refl}$ to the coronal radius ($R_{\rm cor}$ or $R_{\rm warm}$) throughout the analysis, we have in practice assumed a disc/corona geometry where the optically thick plasma is blanketing the accretion disc, and no reflection component is observed from within $R_{\rm cor}$. Now we try to separately determine the region in the disc where the observed \fe line is emitted and the radial size of the warm corona itself. Even when $R_{\rm in,refl}$ is free to vary, although poorly constrained, its value remains much larger than both $R_{\rm isco}$ (irrespective of the black hole spin) and that of the warm/hot corona: $R_{\rm in,refl}=37_{-22}^{+34}\,\rg$ (phenomenological model), $R_{\rm in,refl}=67_{-49}^{+36}\,\rg$ (\optxagn), and $R_{\rm in,refl}>17\,\rg$ (\agnslim). This confirms that the observed reflection component does not originate from the very inner parts of the disc, which is not surprising given the relatively narrow width ($\sigma\sim300$ eV) and symmetric profile of the \fe line.

The possibility that some reflection comes from the inner disc cannot be completely excluded, since the corona will not actually block it. However, Comptonization will add to the stronger relativistic blurring \citep[e.g.,][]{Wilkins15}, likely making any reflection feature hard to disentangle from the the primary continuum. Probing this effect goes beyond the scope of the present work, but this would provide further insights into the properties of the warm/hot corona. 

\subsection{Is there a disc wind in \ton?}
\label{sec:wind}

The likely super-Eddington accretion rate of \ton could be in itself a sufficient ingredient for the launch of an X-ray disc wind \citep[e.g.,][]{Nardini19}, although the viewing angle is also important \citep{Giustini19}. The NLSy\,1 nature of \ton is challenging for the detection of such a wind, though. In fact, the well-known steepness of the X-ray spectra of NLSy\,1's usually implies an insufficient S/N in the 6.7--10 keV band to discover any blueshifted Fe\,K absorption features. However, these appear to be present when very deep exposures are available \citep[e.g.,][]{Parker17}. Alternatively, ultra-fast winds have been detected in NLSy\,1's through the high-resolution grating spectra at soft X-ray energies \citep[e.g.,][]{Longinotti15,Parker17,Kosec18}. 

Unfortunately, during the longest (2015) \xmm observation of \ton, the source was about two times fainter in the soft X-rays than in the (2000) highest state, which was only covered by a short exposure similar to our 2016 observation (when the source was at its lowest). We therefore had to use the stacked spectrum to search for any faint and/or narrow absorption (and emission) features. Only a handful of possible absorption lines were identified through a blind scan, none of which are statistically significant on their own. However, when two of the strongest ones are combined, these are compatible with the main transitions (from O\,\textsc{viii} and C\,\textsc{vi} Ly$\alpha$) in a gas with moderate column density of $\nh<10^{22}$ \cmsq, relatively high ionization of $\logxi \sim 2.7$, and outflowing at $\vout \simeq -0.2c$. After accounting for an absorber with these properties, the fit of the RGS spectrum marginally improves at the 3.1$\sigma$ confidence level. 

No Fe\,K absorption feature has been detected so far in \ton through a standard spectral analysis. Recently, \citet{Igo20} developed a new model-independent method for revealing ultra-fast X-ray outflows in nearby AGN, based on the analysis of the fractional excess variance spectra \citep[e.g.,][]{Parker17PCAIRAS13224,Parker18,Parker20}. In that study, they suggested that \ton hosts an ultra-fast wind with outflow velocity of $\vout = (-0.35^{+0.02}_{-0.05})c$, the most extreme in a sample of 58 bright AGN from the \xmm Science Archive. This result requires substantial caution. The degree of blueshift places the absorption feature at $E \sim 9.7$ keV, close to the high-energy limit of the \xmm EPIC-pn bandpass. Indeed, the relatively low S/N at these energies makes it very difficult to locate any reliable absorption feature in the spectrum. 

While individually not compelling, these indications of X-ray absorption from ionized outflowing gas in \ton certainly deserve further investigation. Future high-quality observations are needed to conclusively establish the presence of signatures from a fast wind in either the soft X-ray or Fe\,K band. Although some key properties of the putative wind, like ionization and, most importantly, column density, are not well constrained at this stage, a solid detection of both the soft X-ray and Fe\,K phases is expected to be easily within reach of a large effective area and high energy resolution mission like \textit{Athena} \citep{Barret18}, and possibly also of the forthcoming \textit{XRISM} \citep{Tashiro18}. The optimal characterization of the soft component would likely require an \textit{Arcus}-like grating specrometre \citep{Smith19} instead.

%\pagebreak
\section{conclusions}
\label{sec:conclusions}

In this paper we have presented a detailed analysis of the 2016 \xmmnu observation of the NLSy\,1 \ton by fitting the EPIC, RGS, OM and FPMA/B spectra. Based on our results, we draw the following conclusions.

\begin{itemize}

	\item We have fitted the X-ray emission in the 0.4--30 keV band with different models of the \relxill family, and we have found that none of the relativistic reflection models alone could self-consistently reproduce the broadband spectrum of \ton. Each of the employed models did provide decent fits on statistical grounds, yet significant residuals in the \fe and hard X-ray bands (especially above $10\kev$) were systematically present. In fact, in attempting to account for the steep and smooth soft excess, parameters such as black hole spin, inner-disc emissivity index and/or source height were forced to extreme values, which, instead, are not required, nor compatible, with the shape and strength of the \fe feature and Compton hump. The reflection fits improved after adding a contribution from a distant neutral reflector (\xillver), despite the lack of a clear narrow \feka core in the spectra. The above residuals, however, were still significant. We also find similar shortcomings with alternative relativistic reflection models from the \textsc{reflkerr} package.

	\item We have analysed in detail the RGS spectrum (stacked over the four 2000--2016 \xmm observations, variable in flux but not in shape) and we find that it is almost completely featureless, consistent with a simple power law of $\Gamma\sim2.9$. This result provided extra support towards the evidence of a soft excess caused by Comptonization rather than blurred reflection as some relativistically broadened emission lines would be still present. None the less, the detection of faint emission from the Ne\,\textsc{ix} triplet suggests that X-ray `bare' AGN have ionized gas outside the line of sight. Tentative evidence is also found for absorption from an accretion disc wind with $\vout \sim -0.2c$, although the statistical significance is limited (3.1$\sigma$).
	
	\item We have then fitted the 2016 optical/UV to hard X-ray SED of \ton with a composite reflection plus phenomenological warm-Comptonization model. In this scenario, the overall 0.4--30 kev emission could be well explained and extrapolated down to a few eV through the contribution of: (i) thermal emission from the outer accretion disc; (ii) Componization of optical/UV `seed' disc-photons from a warm ($kT\sim300$ eV) and optically thick ($\tau\sim10$) plasma, which is supposed to blanket the upper layers of the inner accretion disc; (iii) primary emission dominating above $\sim$1 keV, produced via Compton up-scattering in a standard optically thin ($\tau<0.5$) and hot ($kT\sim100$ keV) corona; (iv) disc reflection responsible for the broad \fe emission at $E\sim6.6$ keV and the mild hard X-ray excess, apparently arising at some distance from the ISCO, where the warm/hot coronae are unlikely to fully cover the accretion disc surface.  
	
	\item We have fitted the 2016 SED of \ton with the physically motivated disc Comptonization model \optxagn. The physical properties of the warm corona are broadly consistent with the phenomenological model and the size of the hot corona was found to be of the order of $\sim$10\,$\rg$. Despite the uncertainty on the black hole mass in \ton, we can conservatively constrain the accretion rate to be at least 50 per cent of Eddington. Considering the high Eddington ratio, for which the assumptions of a standard thin disc might be inadequate, we have also fitted the 2016 SED with the \agnslim model, based on slim disc emissivity developed for super-Eddington accreting sources. The geometrical configuration of \agnslim allows for a radial separation of the hot and warm coronal regions, both of which are confirmed to be rather compact: $R_{\rm hot}\la R_{\rm warm}\la 10\,\rg$. Not surprisingly, this model returned a higher Eddington ratio ($\sim$4--5) for the most likely black hole mass, of the order of $10^{7}\,\Msun$, and a bolometric luminosity, computed between 1 eV and 100 keV, of $L_{\rm bol}\approx 5$--$6\times10^{45}\,\ergs$. This condition is highly favourable to the effective presence of a disc wind. 

\end{itemize}

In a future work, we will investigate whether the broadband spectral/flux changes over the epochs of the four \xmm observations can be ascribed to a change in the physical and geometrical properties of the corona, and we will explore if and how these affect the appearance of the reflection component.

\section{acknowledgements}

We thank the referee for their careful reading and many useful suggestions that helped us improve the clarity of the paper.
This work is based on observations obtained with \xmm, an ESA science mission with instruments and contributions directly funded by ESA Member States and NASA, and with the NuSTAR mission, a project led by the California Institute of Technology, managed by the Jet Propulsion Laboratory, and funded by NASA. This research has made use of the NuSTAR Data Analysis Software (\textsc{nustardas} jointly developed by the ASI Science Data Center and the California Institute of Technology.

GAM, MLP and APL are supported by ESA research fellowships. GAM thank P. O. Petrucci and M. T. Costa for the useful discussion. EN acknowledges financial contribution from the agreement ASI-INAF n.2017-14-H.0 and partial support from the EU Horizon 2020 Marie Sk\l{}odowska-Curie grant agreement no. 664931. JR acknowledges financial support through NASA grant 80NSSC18K1603. VB acknowledges financial support through NASA grant 80NSSC20K0793. RM acknowledges the financial support of INAF (Istituto Nazionale di Astrofisica), Osservatorio Astronomico di Roma, ASI (Agenzia Spaziale Italiana) under contract to INAF: ASI 2014-049-R.0 dedicated to SSDC. WNA acknowledges support from the European Research Council through Advanced Grant 340442, on \textit{Feedback}. MG is supported by the ``Programa de Atracci\'on de Talento'' of the Comunidad de Madrid, grant number 2018-T1/TIC-11733. AMJ acknowledge the National Trainee program and the Irish Research Council.

We acknowledge support from the Faculty of the European Space Astronomy Centre (ESAC) - Funding reference 493. Part of this work was supported by CNES. This research has made use of computing facilities operated by CeSAM data centre at LAM, Marseille, France.

\section*{Data availability}

All the data utilised in this paper are publicly available in the \xmm and \nustar archives. More details of the observations are listed in  Tables\,\ref{tab:summary_obs} and \ref{tab:summary_obs_3epoch}.

\bibliographystyle{mn2e}
\bibliography{tons180_paper_submit_arxiv}

\end{document}